\documentclass[nofootinbib,aps,prl,amsmath,amssymb,reprint,superscriptaddress]{revtex4-2}
\pdfoutput=1
\usepackage[utf8]{inputenc}

\usepackage{graphicx}% Include figure files
\graphicspath{{./figures/}}

\usepackage{tikz}
\usepackage{pgf}
\usepackage{import}

\usepackage{dcolumn}% Align table columns on decimal point

\usepackage{hyperref}

\makeatletter
\let\oldtheequation\theequation
\renewcommand\tagform@[1]{\maketag@@@{\ignorespaces#1\unskip\@@italiccorr}}
\renewcommand\theequation{(\oldtheequation)}
\makeatother
\hypersetup{hidelinks}

\usepackage{amssymb}
\usepackage{mathtools}
\usepackage[thmmarks,amsmath]{ntheorem}

\theoremnumbering{arabic}
\theoremstyle{break}
\theoremsymbol{}
\theoremheaderfont{\normalfont\bfseries}
\theorembodyfont{\upshape}

\newcommand{\comsol}{COMSOL Multiphysics\textsuperscript{\textregistered}}
\newcommand{\clang}{C}
\def\R{\mathbb{R}}
\def\xibold{\boldsymbol{\xi}}

\usepackage{lmodern}

\begin{document}

%\preprint{}

%TC:ignore
\title{Transformation twinning to create isospectral cavities}

% \affiliation command. The \affiliation command should follow the other information
%\affiliation{}
%\thanks{}

\author{Simon~V~Lenz}
\affiliation{School of Biological Sciences, University of Bristol, BS8~1TQ, Bristol, UK}

\author{S\'ebastien~Guenneau}
\affiliation{Department of Mathematics, Imperial College London, SW7~2AZ, London, UK}
\affiliation{UMI 2004 Abraham de Moivre-CNRS, Imperial College London, SW7~2AZ, London, UK}

\author{Bruce~W~Drinkwater}
\affiliation{Department of Mechanical Engineering, University of Bristol, BS8~1TQ, Bristol, UK}

\author{Richard~V~Craster}
\affiliation{Department of Mathematics, Imperial College London, SW7~2AZ, London, UK}
\affiliation{UMI 2004 Abraham de Moivre-CNRS, Imperial College London, SW7~2AZ, London, UK}
\affiliation{Department of Mechanical Engineering, Imperial College London,  SW7~2AZ, London, UK}

\author{Marc~W~Holderied}
\affiliation{School of Biological Sciences, University of Bristol, BS8~1TQ, Bristol, UK}

\date{\today}

\begin{abstract}
Bounded domains have discrete eigenfrequencies/spectra, and cavities with different boundaries and areas have different spectra.
A general methodology for isospectral twinning, whereby the spectra of different cavities are made to coincide, is created by combining ideas from across physics including  transformation optics, inverse problems and metamaterial cloaking.
We twin a hexagonal drum with a deformed hexagonal drum using a non-singular coordinate transform that adjusts the deformed shape by mapping a near boundary domain to a zone of heterogeneous anisotropic medium.
Splines define the mapping zone for twinning these two drums and we verify isospectrality by finite element analysis.
\end{abstract}

\maketitle
%TC:endignore

An open challenge across wave physics is to design cavities that are twins, in the sense we shall define, of a different shaped cavity.
The ability to design such twinned cavities opens up multiple possibilities in say, acoustics, of having two different shaped drums or even rooms/auditoria, sound identical or of two different elastic components sharing the same vibrational eigenfrequencies.
In electromagnetism there are numerous examples of closed cavities \cite{hill2009electromagnetic}, and for water waves having vastly different sized experimental wave tanks share the same eigenfrequencies would be desirable.
In this Letter we create such twinned cavities, the twinning we construct is to ensure that both cavities have identical eigenfrequencies, and as such they are isospectral as they share the same spectrum, despite their different boundaries and areas, and furthermore share the same eigenfields within a well-defined portion of the cavities;
we consider closed cavities for which the spectrum is discrete \cite{hill2009electromagnetic}.
To tackle the challenge of creating twinned cavities we combine ideas from across physics drawing upon transformation optics \cite{Pendry2006,leonhardt2010geometry}, inverse problems \cite{kohn1984determining,greenleaf2003anisotropic} and metamaterial carpet cloaking \cite{li2008hiding}.

Isospectral problems in general have a long history, e.g. in the design of isospectral drums \cite{driscoll2003isospectral,wu1995numerical,driscoll1997eigenmodes} motivated by the  famous question of Mark Kac \cite{kac1966can} as to whether one can hear the shape of a drum.
Kac was referring to the problem of whether the Laplacian operator in a closed domain, with Dirichlet boundary conditions, could have identical spectra on two distinct planar regions sharing the same area.
The question is still open, although there are many results for specific classes and subsets of the problem \cite{gordon1992one}.
These isospectral problems in bounded domains are also related to inverse problems in open space \cite{sleeman1982inverse}.
We use an approach inspired by transformation optics \cite{li2008hiding} that allows us, in contrast to much of the isospectral literature \cite{isozaki2020inverse}, to not limit ourselves to requiring cavities of the same area.
Instead, we modify portions of the original shape to create the isospectral match to the target cavity and in doing so require anisotropic heterogeneous media close to the boundaries.

We shall also draw upon the theory of cloaking \cite{Pendry2006}:
cloaking theories are almost invariably focussed around scattered fields in unbounded domains with line or point source excitation or incoming plane waves and this is quite separate from the isospectral cavity problem.
However, there are very useful concepts that we will utilise:
particularly pertinent is the use of transformation optics in scattering to create mirage \cite{zolla2007electromagnetic} or illusion effects \cite{lai2009illusion,diatta2010non,kan2013acoustic} whereby one object is cloaked such that after scattering it appears to be another object, or where it, or the excitation source, appears to have physically moved.

An important nuance in cloaking revolves around the precise transformation employed and to achieve perfect cloaking a transformation with a singularity is required \cite{Pendry2006}.
This singularity would destroy the discrete nature of the spectrum of the closed cavity should it be employed in the isospectral setting.
However this requirement of extreme material parameters can be relaxed to achieve cloaking over a finite frequency bandwidth by regularising the transform, i.e. blowing up a small ball instead of a point \cite{kohn2008cloaking}.
Another approach that removes the singularity is carpet cloaking \cite{li2008hiding} whereby a curved surface maps onto a flat surface with the objective of hiding an object on a surface.
The approach of \cite{li2008hiding} is attractive in the isospectral setting as it requires no singularity in the transform and yet is tractable to apply.
Hence, we use this transformation idea to deform the boundaries of our target cavity onto those of the reference cavity.
Just as in the carpet cloaking approach this generates a moderately anisotropic and heterogeneous medium.
We are able to employ these ideas for isospectral twinning despite it having no scattering field.

Further, we draw upon the closely related field of inverse problems, \cite{kohn1984determining,lee1989determining}, that aims to uniquely determine some physical parameters, such as an electric conductivity $\sigma$ within a bounded region $\Omega$, by applying a known static voltage $u$ to the surface $\partial\Omega$ and recording the resulting current, $\sigma\nabla u \cdot {\bf n}$, at the boundary %(mathematically a Dirichlet-to-Neumann map)
with ${\bf n}$ as the normal to the surface. Such boundary measurements indeed determine $\sigma$ \cite{kohn2008cloaking}, but only under certain limited conditions:
namely, that $\sigma$ must be scalar-valued, positive and finite.
%However, if some of these conditions are not met (e.g. $\sigma$ is matrix valued) EIT fails  \cite{sylvester1987global}.
%This has been exploited to create non-unique conductivities sharing the same boundary measurements \cite{greenleaf2003anisotropic}.
For the twinned cavities that we design using transformation optics we generate matrix-valued, yet non-singular, conductivities (although our language differs depending upon the physical setting.)
This means that the interior field within the cavity cannot be uniquely determined from the boundary measurements.
In an isospectral problem this would translate into different eigenfields for the cavities and yet both cavities would share the same spectrum.

\begin{figure*}
    %TC:ignore
    \centering
    \includegraphics[width=\textwidth]{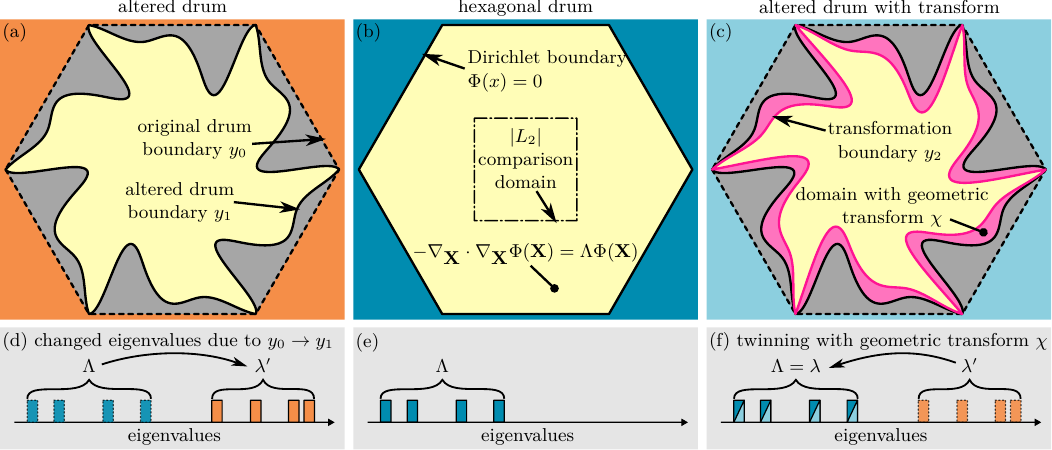}
    %TC:endignore
    \caption{Principle of transformation twinning for the eigenvalues $\Lambda$ and eigenfunctions $\Phi$ in the Laplacian (\autoref{eq:kac}) of a hexagonal drum under Dirichlet boundary conditions. %
    Changing the boundary of a drum (a) with respect to the reference hexagonal drum (b), also results in a change of eigenvalues from $\Lambda$ to $\lambda'$ (d). %
    Using a non-singular coordinate transform (\autoref{eq:carpet}), a perturbed Laplacian with heterogeneous anisotropic parameters (\autoref{eq:prop2}-\ref{eq:prop3}) is introduced into the transformed domain (pink, between $y_1$ and $y_2$) of the altered drum (c). %
    Thus, the eigenvalues $\lambda$ and eigenfunctions $\phi$ outside of the transformed domain (unaltered medium) coincide with those of the original hexagonal drum (e) and the two drums are twinned (f).
    }%
    \label{fig:drum_principle}
\end{figure*}

\begin{figure*}
    %TC:ignore
    \centering
    \includegraphics[width=\textwidth]{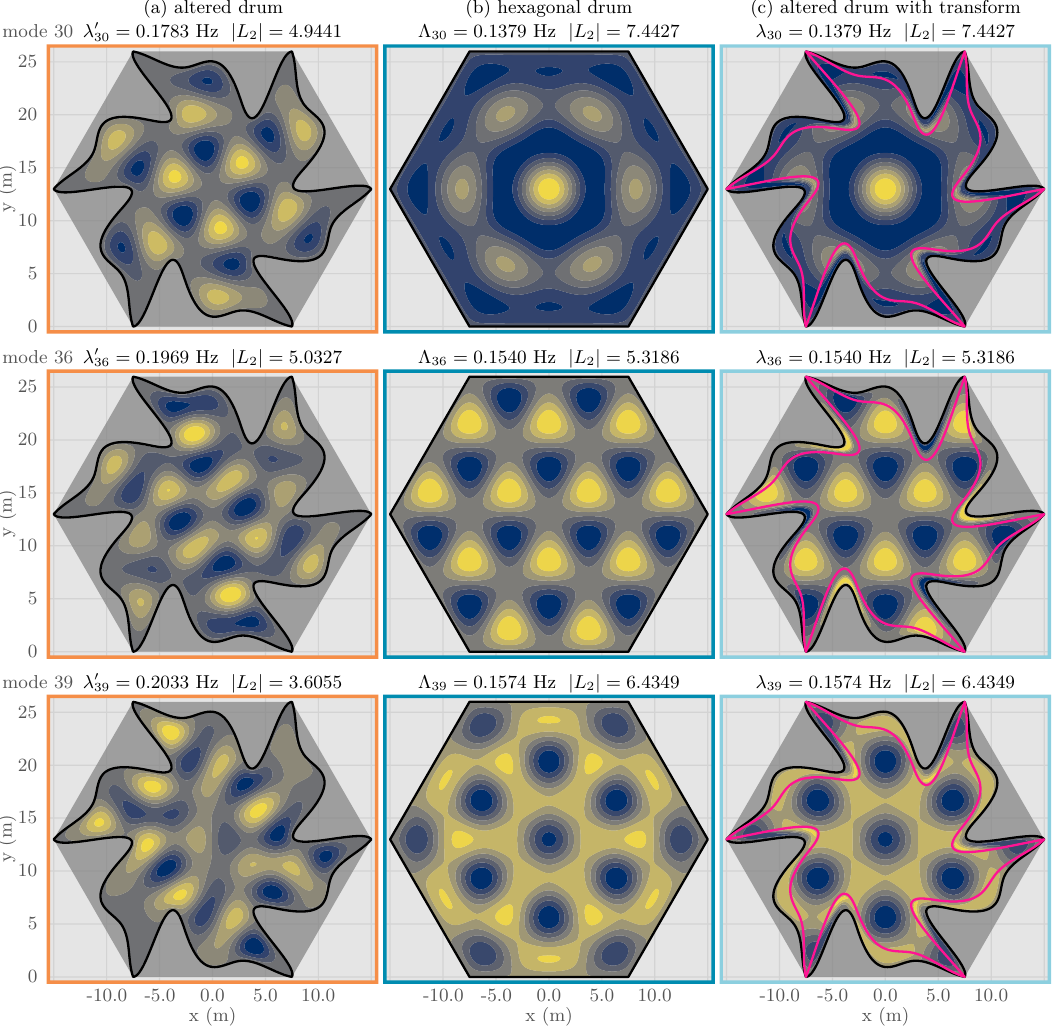}
    %TC:endignore
    \caption{Selection of modes for comparison of three different drums (assuming wave speed $c = 1\;ms^{-1}$, as $\sigma=\eta=1$ in the unaltered medium). %
    The altered drum (a) shows different eigenmodes and eigenfrequencies compared to the hexagonal drum (b), caused by the deformed Dirichlet boundary (black line). %
    By introducing a coordinate transform within a boundary layer (between the black and pink line) of the altered drum (c), the respective eigenfrequencies and eigenmodes coincide with those of the hexagonal drum. %
    Eigenfrequencies and $|L_2|$ are provided above each panel.}%
    \label{fig:drum_compare}
\end{figure*}

\begin{figure}
    %TC:ignore
    \centering
    \includegraphics[width=\columnwidth]{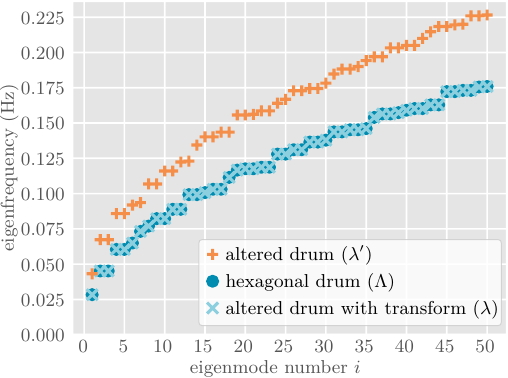}
    %TC:endignore
    \caption{%
      Numerical comparison of the first 50 computed eigenfrequencies (some of which are degenerate) of the hexagonal drum versus the altered and the twinned drum assuming wave speed $c = 1\;ms^{-1}$, as $\sigma=\eta=1$ in the unaltered medium. %
      The eigenfrequencies of the altered drum are generally higher than the hexagonal drum. %
      With the transformation however, the eigenfrequencies of the altered drum match those of the hexagonal drum closely and achieve twinning. %
      The mean absolute difference between twinned and original drum is %
      $3.51e\text{-}8\pm3.50e\text{-}8$~Hz, %
      and $3.85e\text{-}2 \pm 9.37e\text{-}3$~Hz between altered and original drum, %
      for a mesh with 2,500,000 elements (mean $\pm$ standard deviation).%
    }%
    \label{fig:freq_match}
\end{figure}

The challenge is exemplified in \autoref{fig:drum_principle}:
a perfect hexagon (original drum) with the zero (Dirichlet) boundary condition on its boundary ($y_0$) has a discrete spectrum of eigenfrequencies with associated eigenfunctions.
Also shown is a highly deformed ``hexagon" where each edge has been deformed to have boundary $y_1$ (with Dirichlet boundary condition).
The challenge is to have this new shape have the same eigenvalues as the perfect hexagon, this is achieved by a coordinate transformation in a domain close to boundary which effectively squeezes the missing material into the deformed shape via changing the material parameters.
The eigenfield is slightly harder to unequivocally compare between the two shapes, however this is possible by choosing a domain away from the inner boundary and then comparing the eigenfields by an appropriate norm, here we choose the euclidean $L_2$ norm.
We use this hexagon and its deformed counterpart as the exemplar upon which to demonstrate our methodology and, without loss of generality, we restrict our analysis to the two-dimensional case.
\autoref{fig:drum_principle}(d)-(f) also shows, qualitatively, what we want to achieve, the eigenfrequencies for the perfect hexagon are shifted by the shape change and we want to restore them by using the coordinate transformation.

Numerically we adopt a general approach to transformation so we can tackle arbitrary deformations of the boundaries (e.g. splines) that can describe any object/boundary of the shape and this lends our approach versatility.
All of our eigenfunction computations are performed in \comsol{} using standard finite elements to discretise the domains of interest.
The respective transformations are introduced into the finite element simulation via a \clang{} library built from the boundary functions ($y_0$, $y_1$ \& $y_2$).
For our numerical solutions, the eigenfields are normalised to a value range of $[-1, 1]$ with respect to the original drum.
These normalised and reoriented regions are then compared based on their $L_2$ norm (within the centered $8 \times 5$~m comparison domain, see \autoref{fig:drum_principle}(b)), using the numerical integration with Simpson's rule (Newton-Coates order 2) $\sqrt{\int\int u^2 dxdy}$ for the nodal values $u(x, y)$ from the simulation.

In \cite{kac1966can} Kac wondered whether it is possible to deduce the precise shape of a drum just from hearing the fundamental tone and all the overtones? Mathematically this reduces to the study of the eigenvalues, $\Lambda$, and associated eigenfields $\Phi$, of the Laplacian e.g.
%TC:ignore
\begin{equation}
-\Delta_{\textbf{X}}\Phi({\textbf{X}})
=\Lambda \Phi({\textbf{X}})\; \hbox{in a bounded domain} \; \Pi
\label{eq:kac}
\end{equation}
%TC:endignore
where $\Delta_{\textbf{X}}=\partial^2/\partial x^2+\partial^2/\partial y^2$ is the Laplacian in Cartesian coordinates, and where \cite{kac1966can} considered Dirichlet conditions along the boundary $\partial\Pi$ (Neumann conditions also work).
In general, the spectral properties of a shape are linked to its geometry and its material constituent.

Now we introduce a coordinate transformation $\chi$ that maps the cavity of shape $\Pi$, in our example the undeformed hexagon, onto a transformed cavity of shape $\pi$, the deformed hexagon, i.e.
%TC:ignore
\begin{equation}
\chi : \textbf{X}=(x,y)\in\Pi \rightarrow \textbf{s}=(u(x,y),v(x,y))\in\pi
\end{equation}
%TC:endignore
Using the chain rule, we obtain the counterpart of \autoref{eq:kac} in the transformed coordinates as
 %TC:ignore
\begin{equation}
-\nabla_{\textbf{s}}\cdot ( \sigma(\textbf{s}) \nabla_{\textbf{s}} \phi(\textbf{s})) = \eta (\textbf{s}) \lambda \phi(\textbf{s}) \;
\label{eq:prop2}
\end{equation}
%TC:endignore
in the transformed domain $\pi$ where $\nabla_{\textbf{s}}$ is the gradient operator in $u,v$ coordinates, $\phi(\textbf{s}(\textbf{X}))= \Phi(\textbf{X})$ and
%TC:ignore
\begin{equation}
\sigma(\textbf{s}) = {\bf J}_{\bf sX} {\bf J}_{\bf sX}^T \rm{det} {\bf J}_{\bf Xs}
\label{eq:prop3}
\end{equation}
%TC:endignore
is a matrix valued, spatially varying, parameter (physically related to some artificial anisotropic shear modulus, permittivity, permeability or mass density, in anti-plane shear elastic (SH), transverse magnetic (TM), transverse electric (TE) or pressure acoustic settings, respectively), that can be achieved through effective medium theory e.g. with layered media, see Supplemental Material. This parameter $\sigma$ depends on the Jacobian ${\bf J}_{\bf sX}=\partial (u,v)/\partial(x,y)$ of the geometrical transformation $\chi$.
In \autoref{eq:prop2} $\eta = \rm{det} {\bf J}_{\bf Xs}$ is a spatially varying scalar parameter also depending on $\chi$ (physically related to some artificial isotropic mass density, permeability, permittivity, or bulk modulus, in SH, TM, TE or pressure acoustic settings, respectively).
%The coordinate transformations do not affect the boundary conditions and hence
Eigenvalue problem \ref{eq:prop2} is solved subject to the Dirichlet condition, $\phi=0$, on the boundary $\partial\pi$
(or Neumann condition $\sigma\nabla_s \phi \cdot {\bf n}=0$, with {\bf n} the normal to $\partial\pi$, if Neumann datum was assumed for \ref{eq:kac}.)  

As the parameter $\sigma$ in \ref{eq:prop3}, and its counterpart in \ref{eq:prop3rot} are both symmetric, the differential operator in \ref{eq:prop2} is symmetric and its spectrum is real positive.
However, to ensure that its resolvent is compact, and thus that  $0<\lambda_1 \leq \lambda_2 \leq \cdots \leq \lambda_k \cdots$ that tend to $+\infty$, we also require that $\sigma$ be a positive definite and bounded matrix.
In other words, there should exist two positive real constants $m$ and $M$ such that $0<m\leq \sigma \xibold\cdot\xibold\leq M$, for every vector $\xibold$ in $\R^2$.
This criterion is satisfied if the eigenvalues of matrix $\sigma$ are all strictly positive and finite.
This raises an important practical point:
we can only use a class of non-singular transforms to preserve the discrete spectrum, and the mathematical criterion is that the eigenvalues of the Jacobian need to be bounded from below and above by strictly positive constants.
We now consider non-singular geometrical transforms that ensure the resolvent of the differential operator in partial differential equation \ref{eq:prop2} remains compact.

To proceed we borrow the non-singular geometric transforms of Li and Pendry who introduced them for the design of ground carpet cloaks in electromagnetic scattering problems \cite{li2008hiding}.
As shown in \autoref{fig:drum_principle} the region between the outer ($y_2$) and ground boundary ($y_0= 0$) is compressed into the region between outer ($y_2$) and inner ($y_1$) boundary.
We introduce a transformation which maps the region enclosed between two curves $(x,0)$ and $(x,y_2(x))$ to the one comprised between $(x,y_1(x))$ and $(x,y_2(x))$.
This corresponds to a compression of space from the $y_0$-$y_2$ region into the $y_1$-$y_2$ region.
In \autoref{fig:drum_principle}, $(x,0)$ (ground boundary) is mapped on $(x,y_1(x))$ (inner boundary) and  $(x,y_2(x))$ (outer boundary) is fixed point-wise, of the form
%TC:ignore
\begin{equation}\label{eq:carpet}
\left\{
\begin{array}{lr}
u=x\\
v=\alpha(x) y+ \beta(x)
\end{array}
\right.
\end{equation}
%TC:endignore
where
$\alpha (x)= ({y_2-y_1})/{y_2}$ and $\beta(x)= y_1$.
The transformed parameter $\sigma$ is then easily deduced from \ref{eq:prop3} by computing the Jacobian matrix ${\bf J}_{\bf sX}$ associated with \ref{eq:carpet}, and similarly for $\eta$ is deduced from the determinant of the Jacobian.

For our exemplar, the hexagon and its deformed counterpart, we need to do a coordinate transformation where the deformations are identical on each face bar a shift and rotation.
Typically, the transformation concept, as applied to carpet cloaks \cite{li2008hiding} is for flat reference boundaries (see Supplemental Material for curved reference boundaries $y_0\ne 0$), whereas here we have flat walls at certain angles with respect to the horizontal axis.
We thus proceed as follows:
First, we design the transformation for the wall parallel to the $x$-axis, and deduce parameter $\sigma$ for the other five carpets through formula
%TC:ignore
\begin{equation}
{\bf R}({\theta})\sigma(\bf{s}){\bf R}(-{\theta})
\label{eq:prop3rot}
\end{equation}
%TC:endignore
where ${\bf R}({\theta})$ is the rotation matrix through a counter-clockwise angle $\theta$.
In the present case, we consider an hexagonal drum so $\theta=n\pi/3$, with $n=1,..,5$.
We note that parameter $\eta$ in \autoref{eq:prop2} is not affected by the rotation.
One need not assume in general that the $n$ carpet cloaks constituting the polygonal drum are identical, but the $n-1$ carpets associated with walls not parallel to the $x$-axis should be deduced from the transformation design via a rotation of given angle.
We further note that the matrix in \autoref{eq:prop3rot} is symmetric, as required, since $a$ is symmetric and ${\bf R}(-\theta)={\bf R}^T(\theta)$.

In \autoref{fig:drum_compare} we show the results for three modes, with respective eigenfrequencies and $L_2$ norm results.
In the perfect hexagonal drum (\autoref{fig:drum_compare}(b)) the eigenmodes retain the symmetry of the shape and we chose three modes:
one mode with simpler features, one with a rapidly oscillating geometrical mode, and another multipolar mode.
The deformed hexagon (\autoref{fig:drum_compare}(a)), as expected, has dramatically different eigenmodes that bear little relation to that of the perfect hexagon.
The eigenfrequencies are similarly poorly matched to those of the perfect hexagon.
Turning to the deformed hexagon, with the transformation applied, we see the eigenfrequencies coincide with those of the perfect hexagon and, within the non-transformed region (see \autoref{fig:drum_principle}(c)), the $L_2$ norm gives very good agreement:
the cavities are twinned despite the severity of the deformation and the change of area enclosed.
The modes shown are very typical and the accuracy follows across many tens of modes with \autoref{fig:freq_match} showing the first 50 eigenfrequencies.
The presented data corresponds to a frequency range of $0 - 0.23$~Hz, but we can twin the entire spectrum in theory, the only restriction being computational resources and numerical accuracy.
One way towards achieving twinning in an experiment would be the use of a layered medium, with parameters derived from effective medium theory.
We provide a proof-of-concept for this approach in the Supplemental Material.

In this Letter, the recent theories of transformation optics have been applied, not to scattering, but to spectral problems.
In doing so we have shown that we can match the spectrum for strongly different shapes in closed domains, whilst limiting the transformation to a boundary region.
This opens the path to several applications, the spectrum is essential in terms of energy transport for photonic wave\-guides and crystal fibres.
The ideas presented here could be used to retrofit, or repair, waveguides using metamaterial regions of transformed material to create, or recreate, the eigenfrequencies and eigenmodes desired.
Similarly one could envisage cavity devices that, through manufacture or damage, are not operating at the right frequency or which require tunability and for which this transformation methodology would allow eigenfrequencies to be designed for.
In vibration control an example would be shifting the eigenfrequencies of slender bridges to avoid unwanted swaying and resonance effects \cite{brun2013bypassing}.
This also has implications for designed and tunable resonance in particular to decrease a cavity area whilst maintaining the same resonance properties.

On a practical level the effectiveness and quality of cloaking for optics and electromagnetism is hard to quantify and assess, and one reason we were drawn to the identification of the spectrum is that the isospectral problem is a robust, and unequivocal, approach to assess the quality of carpet cloaks (independent of boundary conditions, wave simulations, Perfectly Matched Layers etc.).
The agreement of the twinned spectra is a robust measure of cloaking quality.
The approach here paves the way towards numerous extensions to other wave systems such as the full Maxwell system \cite{nicolet1994transformation,Pendry2006,leonhardt2010geometry,fleury2015invisibility} or to acoustics \cite{popa2011experimental,faure2016experiments} and water waves \cite{berraquero2013experimental,dupont2015numerical} and also to others such as elasticity where perfect cloaking is not available \cite{milton2006cloaking}, but ideas around direct lattice transformation \cite{buckmann2015mechanical} and near-cloaking \cite{QUADRELLI21a} could be adapted, to perforated domains, heterogeneous or anisotropic cavities, and will motivate experiments and devices.

%TC:ignore
\begin{acknowledgements}
The authors are funded by the UK Engineering and Physical Sciences Research Council (EP/T002654/1) and R.V.C. was additionally funded by the H2020 FETOpen project BOHEME under grant agreement No.~863179.
This work was supported by the Biotechnology and Biological Sciences Research Council-funded South West Biosciences Doctoral Training Partnership (BB/M009122/1).
\end{acknowledgements}

\bibliography{bibliography}

\begin{thebibliography}{29}%
\makeatletter
\providecommand \@ifxundefined [1]{%
 \@ifx{#1\undefined}
}%
\providecommand \@ifnum [1]{%
 \ifnum #1\expandafter \@firstoftwo
 \else \expandafter \@secondoftwo
 \fi
}%
\providecommand \@ifx [1]{%
 \ifx #1\expandafter \@firstoftwo
 \else \expandafter \@secondoftwo
 \fi
}%
\providecommand \natexlab [1]{#1}%
\providecommand \enquote  [1]{``#1''}%
\providecommand \bibnamefont  [1]{#1}%
\providecommand \bibfnamefont [1]{#1}%
\providecommand \citenamefont [1]{#1}%
\providecommand \href@noop [0]{\@secondoftwo}%
\providecommand \href [0]{\begingroup \@sanitize@url \@href}%
\providecommand \@href[1]{\@@startlink{#1}\@@href}%
\providecommand \@@href[1]{\endgroup#1\@@endlink}%
\providecommand \@sanitize@url [0]{\catcode `\\12\catcode `\$12\catcode `\&12\catcode `\#12\catcode `\^12\catcode `\_12\catcode `\%12\relax}%
\providecommand \@@startlink[1]{}%
\providecommand \@@endlink[0]{}%
\providecommand \url  [0]{\begingroup\@sanitize@url \@url }%
\providecommand \@url [1]{\endgroup\@href {#1}{\urlprefix }}%
\providecommand \urlprefix  [0]{URL }%
\providecommand \Eprint [0]{\href }%
\providecommand \doibase [0]{https://doi.org/}%
\providecommand \selectlanguage [0]{\@gobble}%
\providecommand \bibinfo  [0]{\@secondoftwo}%
\providecommand \bibfield  [0]{\@secondoftwo}%
\providecommand \translation [1]{[#1]}%
\providecommand \BibitemOpen [0]{}%
\providecommand \bibitemStop [0]{}%
\providecommand \bibitemNoStop [0]{.\EOS\space}%
\providecommand \EOS [0]{\spacefactor3000\relax}%
\providecommand \BibitemShut  [1]{\csname bibitem#1\endcsname}%
\let\auto@bib@innerbib\@empty
%</preamble>
\bibitem [{\citenamefont {Hill}(2009)}]{hill2009electromagnetic}%
  \BibitemOpen
  \bibfield  {author} {\bibinfo {author} {\bibfnamefont {D.~A.}\ \bibnamefont {Hill}},\ }\href@noop {} {\emph {\bibinfo {title} {Electromagnetic fields in cavities: deterministic and statistical theories}}}\ (\bibinfo  {publisher} {John Wiley \& Sons},\ \bibinfo {year} {2009})\BibitemShut {NoStop}%
\bibitem [{\citenamefont {Pendry}\ \emph {et~al.}(2006)\citenamefont {Pendry}, \citenamefont {Schurig},\ and\ \citenamefont {Smith}}]{Pendry2006}%
  \BibitemOpen
  \bibfield  {author} {\bibinfo {author} {\bibfnamefont {J.~B.}\ \bibnamefont {Pendry}}, \bibinfo {author} {\bibfnamefont {D.}~\bibnamefont {Schurig}},\ and\ \bibinfo {author} {\bibfnamefont {D.~R.}\ \bibnamefont {Smith}},\ }\bibfield  {title} {\bibinfo {title} {Controlling electromagnetic fields},\ }\href@noop {} {\bibfield  {journal} {\bibinfo  {journal} {Science}\ }\textbf {\bibinfo {volume} {312}},\ \bibinfo {pages} {1780} (\bibinfo {year} {2006})}\BibitemShut {NoStop}%
\bibitem [{\citenamefont {Leonhardt}\ and\ \citenamefont {Philbin}(2010)}]{leonhardt2010geometry}%
  \BibitemOpen
  \bibfield  {author} {\bibinfo {author} {\bibfnamefont {U.}~\bibnamefont {Leonhardt}}\ and\ \bibinfo {author} {\bibfnamefont {T.}~\bibnamefont {Philbin}},\ }\href@noop {} {\emph {\bibinfo {title} {Geometry and light: the science of invisibility}}}\ (\bibinfo  {publisher} {Courier Corporation},\ \bibinfo {year} {2010})\BibitemShut {NoStop}%
\bibitem [{\citenamefont {Kohn}\ and\ \citenamefont {Vogelius}(1984)}]{kohn1984determining}%
  \BibitemOpen
  \bibfield  {author} {\bibinfo {author} {\bibfnamefont {R.}~\bibnamefont {Kohn}}\ and\ \bibinfo {author} {\bibfnamefont {M.}~\bibnamefont {Vogelius}},\ }\bibfield  {title} {\bibinfo {title} {Determining conductivity by boundary measurements},\ }\href@noop {} {\bibfield  {journal} {\bibinfo  {journal} {Communications on Pure and Applied Mathematics}\ }\textbf {\bibinfo {volume} {37}},\ \bibinfo {pages} {289} (\bibinfo {year} {1984})}\BibitemShut {NoStop}%
\bibitem [{\citenamefont {Greenleaf}\ \emph {et~al.}(2003)\citenamefont {Greenleaf}, \citenamefont {Lassas},\ and\ \citenamefont {Uhlmann}}]{greenleaf2003anisotropic}%
  \BibitemOpen
  \bibfield  {author} {\bibinfo {author} {\bibfnamefont {A.}~\bibnamefont {Greenleaf}}, \bibinfo {author} {\bibfnamefont {M.}~\bibnamefont {Lassas}},\ and\ \bibinfo {author} {\bibfnamefont {G.}~\bibnamefont {Uhlmann}},\ }\bibfield  {title} {\bibinfo {title} {Anisotropic conductivities that cannot be detected by {EIT}},\ }\href@noop {} {\bibfield  {journal} {\bibinfo  {journal} {Physiological measurement}\ }\textbf {\bibinfo {volume} {24}},\ \bibinfo {pages} {413} (\bibinfo {year} {2003})}\BibitemShut {NoStop}%
\bibitem [{\citenamefont {Li}\ and\ \citenamefont {Pendry}(2008)}]{li2008hiding}%
  \BibitemOpen
  \bibfield  {author} {\bibinfo {author} {\bibfnamefont {J.}~\bibnamefont {Li}}\ and\ \bibinfo {author} {\bibfnamefont {J.~B.}\ \bibnamefont {Pendry}},\ }\bibfield  {title} {\bibinfo {title} {Hiding under the carpet: a new strategy for cloaking},\ }\href@noop {} {\bibfield  {journal} {\bibinfo  {journal} {Physical Review Letters}\ }\textbf {\bibinfo {volume} {101}},\ \bibinfo {pages} {203901} (\bibinfo {year} {2008})},\ \Eprint {https://arxiv.org/abs/0806.4396} {0806.4396} \BibitemShut {NoStop}%
\bibitem [{\citenamefont {Driscoll}\ and\ \citenamefont {Gottlieb}(2003)}]{driscoll2003isospectral}%
  \BibitemOpen
  \bibfield  {author} {\bibinfo {author} {\bibfnamefont {T.~A.}\ \bibnamefont {Driscoll}}\ and\ \bibinfo {author} {\bibfnamefont {H.~P.~W.}\ \bibnamefont {Gottlieb}},\ }\bibfield  {title} {\bibinfo {title} {Isospectral shapes with {N}eumann and alternating boundary conditions},\ }\href@noop {} {\bibfield  {journal} {\bibinfo  {journal} {Physical Review E}\ }\textbf {\bibinfo {volume} {68}},\ \bibinfo {pages} {016702} (\bibinfo {year} {2003})}\BibitemShut {NoStop}%
\bibitem [{\citenamefont {Wu}\ \emph {et~al.}(1995)\citenamefont {Wu}, \citenamefont {Sprung},\ and\ \citenamefont {Martorell}}]{wu1995numerical}%
  \BibitemOpen
  \bibfield  {author} {\bibinfo {author} {\bibfnamefont {H.}~\bibnamefont {Wu}}, \bibinfo {author} {\bibfnamefont {D.~W.~L.}\ \bibnamefont {Sprung}},\ and\ \bibinfo {author} {\bibfnamefont {J.}~\bibnamefont {Martorell}},\ }\bibfield  {title} {\bibinfo {title} {Numerical investigation of isospectral cavities built from triangles},\ }\href@noop {} {\bibfield  {journal} {\bibinfo  {journal} {Physical Review E}\ }\textbf {\bibinfo {volume} {51}},\ \bibinfo {pages} {703} (\bibinfo {year} {1995})},\ \Eprint {https://arxiv.org/abs/solv-int/9408001} {solv-int/9408001} \BibitemShut {NoStop}%
\bibitem [{\citenamefont {Driscoll}(1997)}]{driscoll1997eigenmodes}%
  \BibitemOpen
  \bibfield  {author} {\bibinfo {author} {\bibfnamefont {T.~A.}\ \bibnamefont {Driscoll}},\ }\bibfield  {title} {\bibinfo {title} {Eigenmodes of isospectral drums},\ }\href@noop {} {\bibfield  {journal} {\bibinfo  {journal} {SIAM Review}\ }\textbf {\bibinfo {volume} {39}},\ \bibinfo {pages} {1} (\bibinfo {year} {1997})}\BibitemShut {NoStop}%
\bibitem [{\citenamefont {Kac}(1966)}]{kac1966can}%
  \BibitemOpen
  \bibfield  {author} {\bibinfo {author} {\bibfnamefont {M.}~\bibnamefont {Kac}},\ }\bibfield  {title} {\bibinfo {title} {Can one hear the shape of a drum?},\ }\href@noop {} {\bibfield  {journal} {\bibinfo  {journal} {The American Mathematical Monthly}\ }\textbf {\bibinfo {volume} {73}},\ \bibinfo {pages} {1} (\bibinfo {year} {1966})}\BibitemShut {NoStop}%
\bibitem [{\citenamefont {Gordon}\ \emph {et~al.}(1992)\citenamefont {Gordon}, \citenamefont {Webb},\ and\ \citenamefont {Wolpert}}]{gordon1992one}%
  \BibitemOpen
  \bibfield  {author} {\bibinfo {author} {\bibfnamefont {C.}~\bibnamefont {Gordon}}, \bibinfo {author} {\bibfnamefont {D.~L.}\ \bibnamefont {Webb}},\ and\ \bibinfo {author} {\bibfnamefont {S.}~\bibnamefont {Wolpert}},\ }\bibfield  {title} {\bibinfo {title} {One cannot hear the shape of a drum},\ }\href@noop {} {\bibfield  {journal} {\bibinfo  {journal} {Bulletin of the American Mathematical Society}\ }\textbf {\bibinfo {volume} {27}},\ \bibinfo {pages} {134} (\bibinfo {year} {1992})},\ \Eprint {https://arxiv.org/abs/math/9207215} {math/9207215} \BibitemShut {NoStop}%
\bibitem [{\citenamefont {Sleeman}(1982)}]{sleeman1982inverse}%
  \BibitemOpen
  \bibfield  {author} {\bibinfo {author} {\bibfnamefont {B.}~\bibnamefont {Sleeman}},\ }\bibfield  {title} {\bibinfo {title} {The inverse problem of acoustic scattering},\ }\href@noop {} {\bibfield  {journal} {\bibinfo  {journal} {IMA Journal of Applied Mathematics}\ }\textbf {\bibinfo {volume} {29}},\ \bibinfo {pages} {113} (\bibinfo {year} {1982})}\BibitemShut {NoStop}%
\bibitem [{\citenamefont {Isozaki}\ \emph {et~al.}(2020)\citenamefont {Isozaki} \emph {et~al.}}]{isozaki2020inverse}%
  \BibitemOpen
  \bibfield  {author} {\bibinfo {author} {\bibfnamefont {H.}~\bibnamefont {Isozaki}} \emph {et~al.},\ }\href@noop {} {\emph {\bibinfo {title} {Inverse Spectral and Scattering Theory: An Introduction}}}\ (\bibinfo  {publisher} {Springer},\ \bibinfo {year} {2020})\BibitemShut {NoStop}%
\bibitem [{\citenamefont {Zolla}\ \emph {et~al.}(2007)\citenamefont {Zolla}, \citenamefont {Guenneau}, \citenamefont {Nicolet},\ and\ \citenamefont {Pendry}}]{zolla2007electromagnetic}%
  \BibitemOpen
  \bibfield  {author} {\bibinfo {author} {\bibfnamefont {F.}~\bibnamefont {Zolla}}, \bibinfo {author} {\bibfnamefont {S.}~\bibnamefont {Guenneau}}, \bibinfo {author} {\bibfnamefont {A.}~\bibnamefont {Nicolet}},\ and\ \bibinfo {author} {\bibfnamefont {J.}~\bibnamefont {Pendry}},\ }\bibfield  {title} {\bibinfo {title} {Electromagnetic analysis of cylindrical invisibility cloaks and the mirage effect},\ }\href@noop {} {\bibfield  {journal} {\bibinfo  {journal} {Optics Letters}\ }\textbf {\bibinfo {volume} {32}},\ \bibinfo {pages} {1069} (\bibinfo {year} {2007})}\BibitemShut {NoStop}%
\bibitem [{\citenamefont {Lai}\ \emph {et~al.}(2009)\citenamefont {Lai}, \citenamefont {Ng}, \citenamefont {Chen}, \citenamefont {Han}, \citenamefont {Xiao}, \citenamefont {Zhang},\ and\ \citenamefont {Chan}}]{lai2009illusion}%
  \BibitemOpen
  \bibfield  {author} {\bibinfo {author} {\bibfnamefont {Y.}~\bibnamefont {Lai}}, \bibinfo {author} {\bibfnamefont {J.}~\bibnamefont {Ng}}, \bibinfo {author} {\bibfnamefont {H.~Y.}\ \bibnamefont {Chen}}, \bibinfo {author} {\bibfnamefont {D.~Z.}\ \bibnamefont {Han}}, \bibinfo {author} {\bibfnamefont {J.~J.}\ \bibnamefont {Xiao}}, \bibinfo {author} {\bibfnamefont {Z.-Q.}\ \bibnamefont {Zhang}},\ and\ \bibinfo {author} {\bibfnamefont {C.~T.}\ \bibnamefont {Chan}},\ }\bibfield  {title} {\bibinfo {title} {Illusion optics: the optical transformation of an object into another object},\ }\href@noop {} {\bibfield  {journal} {\bibinfo  {journal} {Physical Review Letters}\ }\textbf {\bibinfo {volume} {102}},\ \bibinfo {pages} {253902} (\bibinfo {year} {2009})},\ \Eprint {https://arxiv.org/abs/0905.1484} {0905.1484} \BibitemShut {NoStop}%
\bibitem [{\citenamefont {Diatta}\ and\ \citenamefont {Guenneau}(2010)}]{diatta2010non}%
  \BibitemOpen
  \bibfield  {author} {\bibinfo {author} {\bibfnamefont {A.}~\bibnamefont {Diatta}}\ and\ \bibinfo {author} {\bibfnamefont {S.}~\bibnamefont {Guenneau}},\ }\bibfield  {title} {\bibinfo {title} {Non-singular cloaks allow mimesis},\ }\href@noop {} {\bibfield  {journal} {\bibinfo  {journal} {Journal of Optics}\ }\textbf {\bibinfo {volume} {13}},\ \bibinfo {pages} {024012} (\bibinfo {year} {2010})}\BibitemShut {NoStop}%
\bibitem [{\citenamefont {Kan}\ \emph {et~al.}(2013)\citenamefont {Kan}, \citenamefont {Liang}, \citenamefont {Zhu}, \citenamefont {Li}, \citenamefont {Zou}, \citenamefont {Wu}, \citenamefont {Yang},\ and\ \citenamefont {Cheng}}]{kan2013acoustic}%
  \BibitemOpen
  \bibfield  {author} {\bibinfo {author} {\bibfnamefont {W.}~\bibnamefont {Kan}}, \bibinfo {author} {\bibfnamefont {B.}~\bibnamefont {Liang}}, \bibinfo {author} {\bibfnamefont {X.}~\bibnamefont {Zhu}}, \bibinfo {author} {\bibfnamefont {R.}~\bibnamefont {Li}}, \bibinfo {author} {\bibfnamefont {X.}~\bibnamefont {Zou}}, \bibinfo {author} {\bibfnamefont {H.}~\bibnamefont {Wu}}, \bibinfo {author} {\bibfnamefont {J.}~\bibnamefont {Yang}},\ and\ \bibinfo {author} {\bibfnamefont {J.}~\bibnamefont {Cheng}},\ }\bibfield  {title} {\bibinfo {title} {Acoustic illusion near boundaries of arbitrary curved geometry},\ }\href@noop {} {\bibfield  {journal} {\bibinfo  {journal} {Scientific reports}\ }\textbf {\bibinfo {volume} {3}},\ \bibinfo {pages} {1} (\bibinfo {year} {2013})}\BibitemShut {NoStop}%
\bibitem [{\citenamefont {Kohn}\ \emph {et~al.}(2008)\citenamefont {Kohn}, \citenamefont {Shen}, \citenamefont {Vogelius},\ and\ \citenamefont {Weinstein}}]{kohn2008cloaking}%
  \BibitemOpen
  \bibfield  {author} {\bibinfo {author} {\bibfnamefont {R.~V.}\ \bibnamefont {Kohn}}, \bibinfo {author} {\bibfnamefont {H.}~\bibnamefont {Shen}}, \bibinfo {author} {\bibfnamefont {M.~S.}\ \bibnamefont {Vogelius}},\ and\ \bibinfo {author} {\bibfnamefont {M.~I.}\ \bibnamefont {Weinstein}},\ }\bibfield  {title} {\bibinfo {title} {Cloaking via change of variables in electric impedance tomography},\ }\href@noop {} {\bibfield  {journal} {\bibinfo  {journal} {Inverse Problems}\ }\textbf {\bibinfo {volume} {24}},\ \bibinfo {pages} {015016} (\bibinfo {year} {2008})}\BibitemShut {NoStop}%
\bibitem [{\citenamefont {Lee}\ and\ \citenamefont {Uhlmann}(1989)}]{lee1989determining}%
  \BibitemOpen
  \bibfield  {author} {\bibinfo {author} {\bibfnamefont {J.~M.}\ \bibnamefont {Lee}}\ and\ \bibinfo {author} {\bibfnamefont {G.}~\bibnamefont {Uhlmann}},\ }\bibfield  {title} {\bibinfo {title} {Determining anisotropic real-analytic conductivities by boundary measurements},\ }\href@noop {} {\bibfield  {journal} {\bibinfo  {journal} {Communications on Pure and Applied Mathematics}\ }\textbf {\bibinfo {volume} {42}},\ \bibinfo {pages} {1097} (\bibinfo {year} {1989})}\BibitemShut {NoStop}%
\bibitem [{\citenamefont {Brun}\ \emph {et~al.}(2013)\citenamefont {Brun}, \citenamefont {Movchan}, \citenamefont {Jones},\ and\ \citenamefont {McPhedran}}]{brun2013bypassing}%
  \BibitemOpen
  \bibfield  {author} {\bibinfo {author} {\bibfnamefont {M.}~\bibnamefont {Brun}}, \bibinfo {author} {\bibfnamefont {A.}~\bibnamefont {Movchan}}, \bibinfo {author} {\bibfnamefont {I.}~\bibnamefont {Jones}},\ and\ \bibinfo {author} {\bibfnamefont {R.}~\bibnamefont {McPhedran}},\ }\bibfield  {title} {\bibinfo {title} {Bypassing shake, rattle and roll},\ }\href@noop {} {\bibfield  {journal} {\bibinfo  {journal} {Physics World}\ }\textbf {\bibinfo {volume} {26}},\ \bibinfo {pages} {32} (\bibinfo {year} {2013})}\BibitemShut {NoStop}%
\bibitem [{\citenamefont {Nicolet}\ \emph {et~al.}(1994)\citenamefont {Nicolet}, \citenamefont {Remacle}, \citenamefont {Meys}, \citenamefont {Genon},\ and\ \citenamefont {Legros}}]{nicolet1994transformation}%
  \BibitemOpen
  \bibfield  {author} {\bibinfo {author} {\bibfnamefont {A.}~\bibnamefont {Nicolet}}, \bibinfo {author} {\bibfnamefont {J.-F.}\ \bibnamefont {Remacle}}, \bibinfo {author} {\bibfnamefont {B.}~\bibnamefont {Meys}}, \bibinfo {author} {\bibfnamefont {A.}~\bibnamefont {Genon}},\ and\ \bibinfo {author} {\bibfnamefont {W.}~\bibnamefont {Legros}},\ }\bibfield  {title} {\bibinfo {title} {Transformation methods in computational electromagnetism},\ }\href@noop {} {\bibfield  {journal} {\bibinfo  {journal} {Journal of Applied Physics}\ }\textbf {\bibinfo {volume} {75}},\ \bibinfo {pages} {6036} (\bibinfo {year} {1994})}\BibitemShut {NoStop}%
\bibitem [{\citenamefont {Fleury}\ \emph {et~al.}(2015)\citenamefont {Fleury}, \citenamefont {Monticone},\ and\ \citenamefont {Al{\`u}}}]{fleury2015invisibility}%
  \BibitemOpen
  \bibfield  {author} {\bibinfo {author} {\bibfnamefont {R.}~\bibnamefont {Fleury}}, \bibinfo {author} {\bibfnamefont {F.}~\bibnamefont {Monticone}},\ and\ \bibinfo {author} {\bibfnamefont {A.}~\bibnamefont {Al{\`u}}},\ }\bibfield  {title} {\bibinfo {title} {Invisibility and cloaking: Origins, present, and future perspectives},\ }\href@noop {} {\bibfield  {journal} {\bibinfo  {journal} {Physical Review Applied}\ }\textbf {\bibinfo {volume} {4}},\ \bibinfo {pages} {037001} (\bibinfo {year} {2015})}\BibitemShut {NoStop}%
\bibitem [{\citenamefont {Popa}\ \emph {et~al.}(2011)\citenamefont {Popa}, \citenamefont {Zigoneanu},\ and\ \citenamefont {Cummer}}]{popa2011experimental}%
  \BibitemOpen
  \bibfield  {author} {\bibinfo {author} {\bibfnamefont {B.-I.}\ \bibnamefont {Popa}}, \bibinfo {author} {\bibfnamefont {L.}~\bibnamefont {Zigoneanu}},\ and\ \bibinfo {author} {\bibfnamefont {S.~A.}\ \bibnamefont {Cummer}},\ }\bibfield  {title} {\bibinfo {title} {Experimental acoustic ground cloak in air},\ }\href@noop {} {\bibfield  {journal} {\bibinfo  {journal} {Physical review letters}\ }\textbf {\bibinfo {volume} {106}},\ \bibinfo {pages} {253901} (\bibinfo {year} {2011})}\BibitemShut {NoStop}%
\bibitem [{\citenamefont {Faure}\ \emph {et~al.}(2016)\citenamefont {Faure}, \citenamefont {Richoux}, \citenamefont {F{\'e}lix},\ and\ \citenamefont {Pagneux}}]{faure2016experiments}%
  \BibitemOpen
  \bibfield  {author} {\bibinfo {author} {\bibfnamefont {C.}~\bibnamefont {Faure}}, \bibinfo {author} {\bibfnamefont {O.}~\bibnamefont {Richoux}}, \bibinfo {author} {\bibfnamefont {S.}~\bibnamefont {F{\'e}lix}},\ and\ \bibinfo {author} {\bibfnamefont {V.}~\bibnamefont {Pagneux}},\ }\bibfield  {title} {\bibinfo {title} {Experiments on metasurface carpet cloaking for audible acoustics},\ }\href@noop {} {\bibfield  {journal} {\bibinfo  {journal} {Applied Physics Letters}\ }\textbf {\bibinfo {volume} {108}},\ \bibinfo {pages} {064103} (\bibinfo {year} {2016})}\BibitemShut {NoStop}%
\bibitem [{\citenamefont {Berraquero}\ \emph {et~al.}(2013)\citenamefont {Berraquero}, \citenamefont {Maurel}, \citenamefont {Petitjeans},\ and\ \citenamefont {Pagneux}}]{berraquero2013experimental}%
  \BibitemOpen
  \bibfield  {author} {\bibinfo {author} {\bibfnamefont {C.~P.}\ \bibnamefont {Berraquero}}, \bibinfo {author} {\bibfnamefont {A.}~\bibnamefont {Maurel}}, \bibinfo {author} {\bibfnamefont {P.}~\bibnamefont {Petitjeans}},\ and\ \bibinfo {author} {\bibfnamefont {V.}~\bibnamefont {Pagneux}},\ }\bibfield  {title} {\bibinfo {title} {Experimental realization of a water-wave metamaterial shifter},\ }\href@noop {} {\bibfield  {journal} {\bibinfo  {journal} {Physical Review E}\ }\textbf {\bibinfo {volume} {88}},\ \bibinfo {pages} {051002(R)} (\bibinfo {year} {2013})}\BibitemShut {NoStop}%
\bibitem [{\citenamefont {Dupont}\ \emph {et~al.}(2015)\citenamefont {Dupont}, \citenamefont {Kimmoun}, \citenamefont {Molin}, \citenamefont {Guenneau},\ and\ \citenamefont {Enoch}}]{dupont2015numerical}%
  \BibitemOpen
  \bibfield  {author} {\bibinfo {author} {\bibfnamefont {G.}~\bibnamefont {Dupont}}, \bibinfo {author} {\bibfnamefont {O.}~\bibnamefont {Kimmoun}}, \bibinfo {author} {\bibfnamefont {B.}~\bibnamefont {Molin}}, \bibinfo {author} {\bibfnamefont {S.}~\bibnamefont {Guenneau}},\ and\ \bibinfo {author} {\bibfnamefont {S.}~\bibnamefont {Enoch}},\ }\bibfield  {title} {\bibinfo {title} {Numerical and experimental study of an invisibility carpet in a water channel},\ }\href@noop {} {\bibfield  {journal} {\bibinfo  {journal} {Physical Review E}\ }\textbf {\bibinfo {volume} {91}},\ \bibinfo {pages} {023010} (\bibinfo {year} {2015})}\BibitemShut {NoStop}%
\bibitem [{\citenamefont {Milton}\ \emph {et~al.}(2006)\citenamefont {Milton}, \citenamefont {Briane},\ and\ \citenamefont {Willis}}]{milton2006cloaking}%
  \BibitemOpen
  \bibfield  {author} {\bibinfo {author} {\bibfnamefont {G.~W.}\ \bibnamefont {Milton}}, \bibinfo {author} {\bibfnamefont {M.}~\bibnamefont {Briane}},\ and\ \bibinfo {author} {\bibfnamefont {J.~R.}\ \bibnamefont {Willis}},\ }\bibfield  {title} {\bibinfo {title} {On cloaking for elasticity and physical equations with a transformation invariant form},\ }\href@noop {} {\bibfield  {journal} {\bibinfo  {journal} {New Journal of Physics}\ }\textbf {\bibinfo {volume} {8}},\ \bibinfo {pages} {248} (\bibinfo {year} {2006})}\BibitemShut {NoStop}%
\bibitem [{\citenamefont {B{\"u}ckmann}\ \emph {et~al.}(2015)\citenamefont {B{\"u}ckmann}, \citenamefont {Kadic}, \citenamefont {Schittny},\ and\ \citenamefont {Wegener}}]{buckmann2015mechanical}%
  \BibitemOpen
  \bibfield  {author} {\bibinfo {author} {\bibfnamefont {T.}~\bibnamefont {B{\"u}ckmann}}, \bibinfo {author} {\bibfnamefont {M.}~\bibnamefont {Kadic}}, \bibinfo {author} {\bibfnamefont {R.}~\bibnamefont {Schittny}},\ and\ \bibinfo {author} {\bibfnamefont {M.}~\bibnamefont {Wegener}},\ }\bibfield  {title} {\bibinfo {title} {Mechanical cloak design by direct lattice transformation},\ }\href@noop {} {\bibfield  {journal} {\bibinfo  {journal} {Proceedings of the National Academy of Sciences}\ }\textbf {\bibinfo {volume} {112}},\ \bibinfo {pages} {4930} (\bibinfo {year} {2015})}\BibitemShut {NoStop}%
\bibitem [{\citenamefont {Quadrelli}\ \emph {et~al.}(2021)\citenamefont {Quadrelli}, \citenamefont {Craster}, \citenamefont {Kadic},\ and\ \citenamefont {Braghin}}]{QUADRELLI21a}%
  \BibitemOpen
  \bibfield  {author} {\bibinfo {author} {\bibfnamefont {D.~E.}\ \bibnamefont {Quadrelli}}, \bibinfo {author} {\bibfnamefont {R.}~\bibnamefont {Craster}}, \bibinfo {author} {\bibfnamefont {M.}~\bibnamefont {Kadic}},\ and\ \bibinfo {author} {\bibfnamefont {F.}~\bibnamefont {Braghin}},\ }\bibfield  {title} {\bibinfo {title} {Elastic wave near-cloaking},\ }\href {https://doi.org/https://doi.org/10.1016/j.eml.2021.101262} {\bibfield  {journal} {\bibinfo  {journal} {Extreme Mechanics Letters}\ }\textbf {\bibinfo {volume} {44}},\ \bibinfo {pages} {101262} (\bibinfo {year} {2021})}\BibitemShut {NoStop}%
\end{thebibliography}%


\begin{thebibliography}{21}%
\makeatletter
\providecommand \@ifxundefined [1]{%
 \@ifx{#1\undefined}
}%
\providecommand \@ifnum [1]{%
 \ifnum #1\expandafter \@firstoftwo
 \else \expandafter \@secondoftwo
 \fi
}%
\providecommand \@ifx [1]{%
 \ifx #1\expandafter \@firstoftwo
 \else \expandafter \@secondoftwo
 \fi
}%
\providecommand \natexlab [1]{#1}%
\providecommand \enquote  [1]{``#1''}%
\providecommand \bibnamefont  [1]{#1}%
\providecommand \bibfnamefont [1]{#1}%
\providecommand \citenamefont [1]{#1}%
\providecommand \href@noop [0]{\@secondoftwo}%
\providecommand \href [0]{\begingroup \@sanitize@url \@href}%
\providecommand \@href[1]{\@@startlink{#1}\@@href}%
\providecommand \@@href[1]{\endgroup#1\@@endlink}%
\providecommand \@sanitize@url [0]{\catcode `\\12\catcode `\$12\catcode `\&12\catcode `\#12\catcode `\^12\catcode `\_12\catcode `\%12\relax}%
\providecommand \@@startlink[1]{}%
\providecommand \@@endlink[0]{}%
\providecommand \url  [0]{\begingroup\@sanitize@url \@url }%
\providecommand \@url [1]{\endgroup\@href {#1}{\urlprefix }}%
\providecommand \urlprefix  [0]{URL }%
\providecommand \Eprint [0]{\href }%
\providecommand \doibase [0]{https://doi.org/}%
\providecommand \selectlanguage [0]{\@gobble}%
\providecommand \bibinfo  [0]{\@secondoftwo}%
\providecommand \bibfield  [0]{\@secondoftwo}%
\providecommand \translation [1]{[#1]}%
\providecommand \BibitemOpen [0]{}%
\providecommand \bibitemStop [0]{}%
\providecommand \bibitemNoStop [0]{.\EOS\space}%
\providecommand \EOS [0]{\spacefactor3000\relax}%
\providecommand \BibitemShut  [1]{\csname bibitem#1\endcsname}%
\let\auto@bib@innerbib\@empty
%</preamble>
\bibitem [{\citenamefont {Meurer}\ \emph {et~al.}(2017)\citenamefont {Meurer}, \citenamefont {Smith}, \citenamefont {Paprocki}, \citenamefont {\v{C}ert\'{i}k}, \citenamefont {Kirpichev}, \citenamefont {Rocklin}, \citenamefont {Kumar}, \citenamefont {Ivanov}, \citenamefont {Moore}, \citenamefont {Singh}, \citenamefont {Rathnayake}, \citenamefont {Vig}, \citenamefont {Granger}, \citenamefont {Muller}, \citenamefont {Bonazzi}, \citenamefont {Gupta}, \citenamefont {Vats}, \citenamefont {Johansson}, \citenamefont {Pedregosa}, \citenamefont {Curry}, \citenamefont {Terrel}, \citenamefont {Rou\v{c}ka}, \citenamefont {Saboo}, \citenamefont {Fernando}, \citenamefont {Kulal}, \citenamefont {Cimrman},\ and\ \citenamefont {Scopatz}}]{sympy}%
  \BibitemOpen
  \bibfield  {author} {\bibinfo {author} {\bibfnamefont {A.}~\bibnamefont {Meurer}}, \bibinfo {author} {\bibfnamefont {C.~P.}\ \bibnamefont {Smith}}, \bibinfo {author} {\bibfnamefont {M.}~\bibnamefont {Paprocki}}, \bibinfo {author} {\bibfnamefont {O.}~\bibnamefont {\v{C}ert\'{i}k}}, \bibinfo {author} {\bibfnamefont {S.~B.}\ \bibnamefont {Kirpichev}}, \bibinfo {author} {\bibfnamefont {M.}~\bibnamefont {Rocklin}}, \bibinfo {author} {\bibfnamefont {A.}~\bibnamefont {Kumar}}, \bibinfo {author} {\bibfnamefont {S.}~\bibnamefont {Ivanov}}, \bibinfo {author} {\bibfnamefont {J.~K.}\ \bibnamefont {Moore}}, \bibinfo {author} {\bibfnamefont {S.}~\bibnamefont {Singh}}, \bibinfo {author} {\bibfnamefont {T.}~\bibnamefont {Rathnayake}}, \bibinfo {author} {\bibfnamefont {S.}~\bibnamefont {Vig}}, \bibinfo {author} {\bibfnamefont {B.~E.}\ \bibnamefont {Granger}}, \bibinfo {author} {\bibfnamefont {R.~P.}\ \bibnamefont {Muller}}, \bibinfo {author} {\bibfnamefont {F.}~\bibnamefont {Bonazzi}}, \bibinfo {author} {\bibfnamefont {H.}~\bibnamefont {Gupta}}, \bibinfo {author} {\bibfnamefont {S.}~\bibnamefont {Vats}}, \bibinfo {author} {\bibfnamefont {F.}~\bibnamefont {Johansson}}, \bibinfo {author} {\bibfnamefont {F.}~\bibnamefont {Pedregosa}}, \bibinfo {author} {\bibfnamefont {M.~J.}\ \bibnamefont {Curry}}, \bibinfo {author} {\bibfnamefont {A.~R.}\ \bibnamefont {Terrel}}, \bibinfo {author} {\bibfnamefont {v.}~\bibnamefont {Rou\v{c}ka}}, \bibinfo {author} {\bibfnamefont {A.}~\bibnamefont {Saboo}}, \bibinfo {author} {\bibfnamefont {I.}~\bibnamefont {Fernando}}, \bibinfo {author} {\bibfnamefont {S.}~\bibnamefont {Kulal}}, \bibinfo {author} {\bibfnamefont {R.}~\bibnamefont {Cimrman}},\ and\ \bibinfo {author} {\bibfnamefont {A.}~\bibnamefont {Scopatz}},\ }\bibfield  {title} {\bibinfo {title} {Sympy: symbolic computing in python},\ }\href {https://doi.org/10.7717/peerj-cs.103} {\bibfield  {journal} {\bibinfo  {journal} {PeerJ Computer Science}\ }\textbf {\bibinfo {volume} {3}},\ \bibinfo {pages}
  {e103} (\bibinfo {year} {2017})}\BibitemShut {NoStop}%
\bibitem [{\citenamefont {Harris}\ \emph {et~al.}(2020)\citenamefont {Harris}, \citenamefont {Millman}, \citenamefont {van~der Walt}, \citenamefont {Gommers}, \citenamefont {Virtanen}, \citenamefont {Cournapeau}, \citenamefont {Wieser}, \citenamefont {Taylor}, \citenamefont {Berg}, \citenamefont {Smith}, \citenamefont {Kern}, \citenamefont {Picus}, \citenamefont {Hoyer}, \citenamefont {van Kerkwijk}, \citenamefont {Brett}, \citenamefont {Haldane}, \citenamefont {del R{\'{i}}o}, \citenamefont {Wiebe}, \citenamefont {Peterson}, \citenamefont {G{\'{e}}rard-Marchant}, \citenamefont {Sheppard}, \citenamefont {Reddy}, \citenamefont {Weckesser}, \citenamefont {Abbasi}, \citenamefont {Gohlke},\ and\ \citenamefont {Oliphant}}]{numpy}%
  \BibitemOpen
  \bibfield  {author} {\bibinfo {author} {\bibfnamefont {C.~R.}\ \bibnamefont {Harris}}, \bibinfo {author} {\bibfnamefont {K.~J.}\ \bibnamefont {Millman}}, \bibinfo {author} {\bibfnamefont {S.~J.}\ \bibnamefont {van~der Walt}}, \bibinfo {author} {\bibfnamefont {R.}~\bibnamefont {Gommers}}, \bibinfo {author} {\bibfnamefont {P.}~\bibnamefont {Virtanen}}, \bibinfo {author} {\bibfnamefont {D.}~\bibnamefont {Cournapeau}}, \bibinfo {author} {\bibfnamefont {E.}~\bibnamefont {Wieser}}, \bibinfo {author} {\bibfnamefont {J.}~\bibnamefont {Taylor}}, \bibinfo {author} {\bibfnamefont {S.}~\bibnamefont {Berg}}, \bibinfo {author} {\bibfnamefont {N.~J.}\ \bibnamefont {Smith}}, \bibinfo {author} {\bibfnamefont {R.}~\bibnamefont {Kern}}, \bibinfo {author} {\bibfnamefont {M.}~\bibnamefont {Picus}}, \bibinfo {author} {\bibfnamefont {S.}~\bibnamefont {Hoyer}}, \bibinfo {author} {\bibfnamefont {M.~H.}\ \bibnamefont {van Kerkwijk}}, \bibinfo {author} {\bibfnamefont {M.}~\bibnamefont {Brett}}, \bibinfo {author} {\bibfnamefont {A.}~\bibnamefont {Haldane}}, \bibinfo {author} {\bibfnamefont {J.~F.}\ \bibnamefont {del R{\'{i}}o}}, \bibinfo {author} {\bibfnamefont {M.}~\bibnamefont {Wiebe}}, \bibinfo {author} {\bibfnamefont {P.}~\bibnamefont {Peterson}}, \bibinfo {author} {\bibfnamefont {P.}~\bibnamefont {G{\'{e}}rard-Marchant}}, \bibinfo {author} {\bibfnamefont {K.}~\bibnamefont {Sheppard}}, \bibinfo {author} {\bibfnamefont {T.}~\bibnamefont {Reddy}}, \bibinfo {author} {\bibfnamefont {W.}~\bibnamefont {Weckesser}}, \bibinfo {author} {\bibfnamefont {H.}~\bibnamefont {Abbasi}}, \bibinfo {author} {\bibfnamefont {C.}~\bibnamefont {Gohlke}},\ and\ \bibinfo {author} {\bibfnamefont {T.~E.}\ \bibnamefont {Oliphant}},\ }\bibfield  {title} {\bibinfo {title} {Array programming with {NumPy}},\ }\href {https://doi.org/10.1038/s41586-020-2649-2} {\bibfield  {journal} {\bibinfo  {journal} {Nature}\ }\textbf {\bibinfo {volume} {585}},\ \bibinfo {pages} {357} (\bibinfo {year} {2020})}\BibitemShut {NoStop}%
\bibitem [{\citenamefont {Virtanen}\ \emph {et~al.}(2020)\citenamefont {Virtanen}, \citenamefont {Gommers}, \citenamefont {Oliphant}, \citenamefont {Haberland}, \citenamefont {Reddy}, \citenamefont {Cournapeau}, \citenamefont {Burovski}, \citenamefont {Peterson}, \citenamefont {Weckesser}, \citenamefont {Bright}, \citenamefont {van~der Walt}, \citenamefont {Brett}, \citenamefont {Wilson}, \citenamefont {Millman}, \citenamefont {Mayorov}, \citenamefont {Nelson}, \citenamefont {Jones}, \citenamefont {Kern}, \citenamefont {Larson}, \citenamefont {Carey}, \citenamefont {Polat}, \citenamefont {Feng}, \citenamefont {Moore}, \citenamefont {VanderPlas}, \citenamefont {Laxalde}, \citenamefont {Perktold}, \citenamefont {Cimrman}, \citenamefont {Henriksen}, \citenamefont {Quintero}, \citenamefont {Harris}, \citenamefont {Archibald}, \citenamefont {Ribeiro}, \citenamefont {Pedregosa}, \citenamefont {van Mulbregt},\ and\ \citenamefont {{SciPy 1.0 Contributors}}}]{scipy}%
  \BibitemOpen
  \bibfield  {author} {\bibinfo {author} {\bibfnamefont {P.}~\bibnamefont {Virtanen}}, \bibinfo {author} {\bibfnamefont {R.}~\bibnamefont {Gommers}}, \bibinfo {author} {\bibfnamefont {T.~E.}\ \bibnamefont {Oliphant}}, \bibinfo {author} {\bibfnamefont {M.}~\bibnamefont {Haberland}}, \bibinfo {author} {\bibfnamefont {T.}~\bibnamefont {Reddy}}, \bibinfo {author} {\bibfnamefont {D.}~\bibnamefont {Cournapeau}}, \bibinfo {author} {\bibfnamefont {E.}~\bibnamefont {Burovski}}, \bibinfo {author} {\bibfnamefont {P.}~\bibnamefont {Peterson}}, \bibinfo {author} {\bibfnamefont {W.}~\bibnamefont {Weckesser}}, \bibinfo {author} {\bibfnamefont {J.}~\bibnamefont {Bright}}, \bibinfo {author} {\bibfnamefont {S.~J.}\ \bibnamefont {van~der Walt}}, \bibinfo {author} {\bibfnamefont {M.}~\bibnamefont {Brett}}, \bibinfo {author} {\bibfnamefont {J.}~\bibnamefont {Wilson}}, \bibinfo {author} {\bibfnamefont {K.~J.}\ \bibnamefont {Millman}}, \bibinfo {author} {\bibfnamefont {N.}~\bibnamefont {Mayorov}}, \bibinfo {author} {\bibfnamefont {A.~R.~J.}\ \bibnamefont {Nelson}}, \bibinfo {author} {\bibfnamefont {E.}~\bibnamefont {Jones}}, \bibinfo {author} {\bibfnamefont {R.}~\bibnamefont {Kern}}, \bibinfo {author} {\bibfnamefont {E.}~\bibnamefont {Larson}}, \bibinfo {author} {\bibfnamefont {C.~J.}\ \bibnamefont {Carey}}, \bibinfo {author} {\bibfnamefont {{\. I}.}~\bibnamefont {Polat}}, \bibinfo {author} {\bibfnamefont {Y.}~\bibnamefont {Feng}}, \bibinfo {author} {\bibfnamefont {E.~W.}\ \bibnamefont {Moore}}, \bibinfo {author} {\bibfnamefont {J.}~\bibnamefont {VanderPlas}}, \bibinfo {author} {\bibfnamefont {D.}~\bibnamefont {Laxalde}}, \bibinfo {author} {\bibfnamefont {J.}~\bibnamefont {Perktold}}, \bibinfo {author} {\bibfnamefont {R.}~\bibnamefont {Cimrman}}, \bibinfo {author} {\bibfnamefont {I.}~\bibnamefont {Henriksen}}, \bibinfo {author} {\bibfnamefont {E.~A.}\ \bibnamefont {Quintero}}, \bibinfo {author} {\bibfnamefont {C.~R.}\ \bibnamefont {Harris}}, \bibinfo {author} {\bibfnamefont {A.~M.}\
  \bibnamefont {Archibald}}, \bibinfo {author} {\bibfnamefont {A.~H.}\ \bibnamefont {Ribeiro}}, \bibinfo {author} {\bibfnamefont {F.}~\bibnamefont {Pedregosa}}, \bibinfo {author} {\bibfnamefont {P.}~\bibnamefont {van Mulbregt}},\ and\ \bibinfo {author} {\bibnamefont {{SciPy 1.0 Contributors}}},\ }\bibfield  {title} {\bibinfo {title} {{{SciPy} 1.0: Fundamental Algorithms for Scientific Computing in Python}},\ }\href {https://doi.org/10.1038/s41592-019-0686-2} {\bibfield  {journal} {\bibinfo  {journal} {Nature Methods}\ }\textbf {\bibinfo {volume} {17}},\ \bibinfo {pages} {261} (\bibinfo {year} {2020})}\BibitemShut {NoStop}%
\bibitem [{\citenamefont {Hunter}(2007)}]{matplotlib}%
  \BibitemOpen
  \bibfield  {author} {\bibinfo {author} {\bibfnamefont {J.~D.}\ \bibnamefont {Hunter}},\ }\bibfield  {title} {\bibinfo {title} {Matplotlib: A 2d graphics environment},\ }\href {https://doi.org/10.1109/MCSE.2007.55} {\bibfield  {journal} {\bibinfo  {journal} {Computing in Science Engineering}\ }\textbf {\bibinfo {volume} {9}},\ \bibinfo {pages} {90} (\bibinfo {year} {2007})}\BibitemShut {NoStop}%
\bibitem [{\citenamefont {O'Brien}\ and\ \citenamefont {Pendry}(2002)}]{o2002photonic}%
  \BibitemOpen
  \bibfield  {author} {\bibinfo {author} {\bibfnamefont {S.}~\bibnamefont {O'Brien}}\ and\ \bibinfo {author} {\bibfnamefont {J.~B.}\ \bibnamefont {Pendry}},\ }\bibfield  {title} {\bibinfo {title} {Photonic band-gap effects and magnetic activity in dielectric composites},\ }\href@noop {} {\bibfield  {journal} {\bibinfo  {journal} {Journal of Physics: Condensed Matter}\ }\textbf {\bibinfo {volume} {14}},\ \bibinfo {pages} {4035} (\bibinfo {year} {2002})}\BibitemShut {NoStop}%
\bibitem [{\citenamefont {Bensoussan}\ \emph {et~al.}(1978)\citenamefont {Bensoussan}, \citenamefont {Lions},\ and\ \citenamefont {Papanicolaou}}]{Bensoussan1978}%
  \BibitemOpen
  \bibfield  {author} {\bibinfo {author} {\bibfnamefont {A.}~\bibnamefont {Bensoussan}}, \bibinfo {author} {\bibfnamefont {J.-L.}\ \bibnamefont {Lions}},\ and\ \bibinfo {author} {\bibfnamefont {G.}~\bibnamefont {Papanicolaou}},\ }\href@noop {} {\emph {\bibinfo {title} {Asymptotic {Analysis} for {Periodic} {Structures}}}}\ (\bibinfo  {publisher} {American Mathematical Soc.},\ \bibinfo {year} {1978})\BibitemShut {NoStop}%
\bibitem [{\citenamefont {Milton}(2002)}]{Milton2002}%
  \BibitemOpen
  \bibfield  {author} {\bibinfo {author} {\bibfnamefont {G.~W.}\ \bibnamefont {Milton}},\ }\href@noop {} {\emph {\bibinfo {title} {The {Theory} of {Composites}}}}\ (\bibinfo  {publisher} {Cambridge University Press},\ \bibinfo {year} {2002})\BibitemShut {NoStop}%
\bibitem [{\citenamefont {Petiteau}\ \emph {et~al.}(2014)\citenamefont {Petiteau}, \citenamefont {Guenneau}, \citenamefont {Bellieud}, \citenamefont {Zerrad},\ and\ \citenamefont {Amra}}]{Petiteau2014}%
  \BibitemOpen
  \bibfield  {author} {\bibinfo {author} {\bibfnamefont {D.}~\bibnamefont {Petiteau}}, \bibinfo {author} {\bibfnamefont {S.}~\bibnamefont {Guenneau}}, \bibinfo {author} {\bibfnamefont {M.}~\bibnamefont {Bellieud}}, \bibinfo {author} {\bibfnamefont {M.}~\bibnamefont {Zerrad}},\ and\ \bibinfo {author} {\bibfnamefont {C.}~\bibnamefont {Amra}},\ }\bibfield  {title} {\bibinfo {title} {Spectral effectiveness of engineered thermal cloaks in the frequency regime},\ }\href {https://doi.org/10.1038/srep07386} {\bibfield  {journal} {\bibinfo  {journal} {Scientific Reports}\ }\textbf {\bibinfo {volume} {4}},\ \bibinfo {pages} {7386} (\bibinfo {year} {2014})}\BibitemShut {NoStop}%
\bibitem [{\citenamefont {Pomot}\ \emph {et~al.}(2020)\citenamefont {Pomot}, \citenamefont {Payan}, \citenamefont {Remillieux},\ and\ \citenamefont {Guenneau}}]{pomot2020acoustic}%
  \BibitemOpen
  \bibfield  {author} {\bibinfo {author} {\bibfnamefont {L.}~\bibnamefont {Pomot}}, \bibinfo {author} {\bibfnamefont {C.}~\bibnamefont {Payan}}, \bibinfo {author} {\bibfnamefont {M.}~\bibnamefont {Remillieux}},\ and\ \bibinfo {author} {\bibfnamefont {S.}~\bibnamefont {Guenneau}},\ }\bibfield  {title} {\bibinfo {title} {Acoustic cloaking: Geometric transform, homogenization and a genetic algorithm},\ }\href@noop {} {\bibfield  {journal} {\bibinfo  {journal} {Wave Motion}\ }\textbf {\bibinfo {volume} {92}},\ \bibinfo {pages} {102413} (\bibinfo {year} {2020})}\BibitemShut {NoStop}%
\bibitem [{\citenamefont {Guenneau}\ and\ \citenamefont {Zolla}(2000)}]{guenneau2000homogenization}%
  \BibitemOpen
  \bibfield  {author} {\bibinfo {author} {\bibfnamefont {S.}~\bibnamefont {Guenneau}}\ and\ \bibinfo {author} {\bibfnamefont {F.}~\bibnamefont {Zolla}},\ }\bibfield  {title} {\bibinfo {title} {Homogenization of three-dimensional finite photonic crystals},\ }\href@noop {} {\bibfield  {journal} {\bibinfo  {journal} {Progress in Electromagnetics Research}\ }\textbf {\bibinfo {volume} {27}},\ \bibinfo {pages} {91} (\bibinfo {year} {2000})}\BibitemShut {NoStop}%
\bibitem [{\citenamefont {Kristensson}\ and\ \citenamefont {Wellander}(2003)}]{kristensson2003homogenization}%
  \BibitemOpen
  \bibfield  {author} {\bibinfo {author} {\bibfnamefont {G.}~\bibnamefont {Kristensson}}\ and\ \bibinfo {author} {\bibfnamefont {N.}~\bibnamefont {Wellander}},\ }\bibfield  {title} {\bibinfo {title} {Homogenization of the maxwell equations at fixed frequency},\ }\href@noop {} {\bibfield  {journal} {\bibinfo  {journal} {SIAM Journal on Applied Mathematics}\ }\textbf {\bibinfo {volume} {64}},\ \bibinfo {pages} {170} (\bibinfo {year} {2003})}\BibitemShut {NoStop}%
\bibitem [{\citenamefont {Guenneau}\ \emph {et~al.}(2007)\citenamefont {Guenneau}, \citenamefont {Zolla},\ and\ \citenamefont {Nicolet}}]{guenneau2007homogenization}%
  \BibitemOpen
  \bibfield  {author} {\bibinfo {author} {\bibfnamefont {S.}~\bibnamefont {Guenneau}}, \bibinfo {author} {\bibfnamefont {F.}~\bibnamefont {Zolla}},\ and\ \bibinfo {author} {\bibfnamefont {A.}~\bibnamefont {Nicolet}},\ }\bibfield  {title} {\bibinfo {title} {Homogenization of 3d finite photonic crystals with heterogeneous permittivity and permeability},\ }\href@noop {} {\bibfield  {journal} {\bibinfo  {journal} {Waves in Random and Complex Media}\ }\textbf {\bibinfo {volume} {17}},\ \bibinfo {pages} {653} (\bibinfo {year} {2007})}\BibitemShut {NoStop}%
\bibitem [{\citenamefont {Bossavit}(1990)}]{bossavit1990solving}%
  \BibitemOpen
  \bibfield  {author} {\bibinfo {author} {\bibfnamefont {A.}~\bibnamefont {Bossavit}},\ }\bibfield  {title} {\bibinfo {title} {Solving maxwell equations in a closed cavity, and the question of'spurious modes'},\ }\href@noop {} {\bibfield  {journal} {\bibinfo  {journal} {IEEE Transactions on magnetics}\ }\textbf {\bibinfo {volume} {26}},\ \bibinfo {pages} {702} (\bibinfo {year} {1990})}\BibitemShut {NoStop}%
\bibitem [{\citenamefont {Zolla}\ \emph {et~al.}(2005)\citenamefont {Zolla}, \citenamefont {Renversez}, \citenamefont {Nicolet}, \citenamefont {Kuhlmey}, \citenamefont {Guenneau},\ and\ \citenamefont {Felbacq}}]{zolla2005foundations}%
  \BibitemOpen
  \bibfield  {author} {\bibinfo {author} {\bibfnamefont {F.}~\bibnamefont {Zolla}}, \bibinfo {author} {\bibfnamefont {G.}~\bibnamefont {Renversez}}, \bibinfo {author} {\bibfnamefont {A.}~\bibnamefont {Nicolet}}, \bibinfo {author} {\bibfnamefont {B.}~\bibnamefont {Kuhlmey}}, \bibinfo {author} {\bibfnamefont {S.~R.}\ \bibnamefont {Guenneau}},\ and\ \bibinfo {author} {\bibfnamefont {D.}~\bibnamefont {Felbacq}},\ }\href@noop {} {\emph {\bibinfo {title} {Foundations of photonic crystal fibres}}}\ (\bibinfo  {publisher} {World Scientific},\ \bibinfo {year} {2005})\BibitemShut {NoStop}%
\bibitem [{\citenamefont {Monk}(2003)}]{monk2003finite}%
  \BibitemOpen
  \bibfield  {author} {\bibinfo {author} {\bibfnamefont {P.}~\bibnamefont {Monk}},\ }\href@noop {} {\emph {\bibinfo {title} {Finite element methods for Maxwell's equations}}}\ (\bibinfo  {publisher} {Oxford University Press},\ \bibinfo {year} {2003})\BibitemShut {NoStop}%
\bibitem [{\citenamefont {Craster}\ \emph {et~al.}(2010)\citenamefont {Craster}, \citenamefont {Kaplunov},\ and\ \citenamefont {Pichugin}}]{craster2010high}%
  \BibitemOpen
  \bibfield  {author} {\bibinfo {author} {\bibfnamefont {R.~V.}\ \bibnamefont {Craster}}, \bibinfo {author} {\bibfnamefont {J.}~\bibnamefont {Kaplunov}},\ and\ \bibinfo {author} {\bibfnamefont {A.~V.}\ \bibnamefont {Pichugin}},\ }\bibfield  {title} {\bibinfo {title} {High-frequency homogenization for periodic media},\ }\href@noop {} {\bibfield  {journal} {\bibinfo  {journal} {Proceedings of the Royal Society A: Mathematical, Physical and Engineering Sciences}\ }\textbf {\bibinfo {volume} {466}},\ \bibinfo {pages} {2341} (\bibinfo {year} {2010})}\BibitemShut {NoStop}%
\bibitem [{\citenamefont {Allaire}\ and\ \citenamefont {Conca}(1998)}]{Allaire1998}%
  \BibitemOpen
  \bibfield  {author} {\bibinfo {author} {\bibfnamefont {G.}~\bibnamefont {Allaire}}\ and\ \bibinfo {author} {\bibfnamefont {C.}~\bibnamefont {Conca}},\ }\bibfield  {title} {\bibinfo {title} {Bloch wave homogenization and spectral asymptotic analysis},\ }\href@noop {} {\bibfield  {journal} {\bibinfo  {journal} {J. Math. Pure Appl.}\ }\textbf {\bibinfo {volume} {77}},\ \bibinfo {pages} {153} (\bibinfo {year} {1998})}\BibitemShut {NoStop}%
\bibitem [{\citenamefont {Popa}\ \emph {et~al.}(2011)\citenamefont {Popa}, \citenamefont {Zigoneanu},\ and\ \citenamefont {Cummer}}]{popa2011experimental}%
  \BibitemOpen
  \bibfield  {author} {\bibinfo {author} {\bibfnamefont {B.-I.}\ \bibnamefont {Popa}}, \bibinfo {author} {\bibfnamefont {L.}~\bibnamefont {Zigoneanu}},\ and\ \bibinfo {author} {\bibfnamefont {S.~A.}\ \bibnamefont {Cummer}},\ }\bibfield  {title} {\bibinfo {title} {Experimental acoustic ground cloak in air},\ }\href@noop {} {\bibfield  {journal} {\bibinfo  {journal} {Physical review letters}\ }\textbf {\bibinfo {volume} {106}},\ \bibinfo {pages} {253901} (\bibinfo {year} {2011})}\BibitemShut {NoStop}%
\bibitem [{\citenamefont {Ergin}\ \emph {et~al.}(2010)\citenamefont {Ergin}, \citenamefont {Stenger}, \citenamefont {Brenner}, \citenamefont {Pendry},\ and\ \citenamefont {Wegener}}]{ergin2010three}%
  \BibitemOpen
  \bibfield  {author} {\bibinfo {author} {\bibfnamefont {T.}~\bibnamefont {Ergin}}, \bibinfo {author} {\bibfnamefont {N.}~\bibnamefont {Stenger}}, \bibinfo {author} {\bibfnamefont {P.}~\bibnamefont {Brenner}}, \bibinfo {author} {\bibfnamefont {J.~B.}\ \bibnamefont {Pendry}},\ and\ \bibinfo {author} {\bibfnamefont {M.}~\bibnamefont {Wegener}},\ }\bibfield  {title} {\bibinfo {title} {Three-dimensional invisibility cloak at optical wavelengths},\ }\href@noop {} {\bibfield  {journal} {\bibinfo  {journal} {science}\ }\textbf {\bibinfo {volume} {328}},\ \bibinfo {pages} {337} (\bibinfo {year} {2010})}\BibitemShut {NoStop}%
\bibitem [{\citenamefont {Berenger}(1996)}]{berenger1996perfectly}%
  \BibitemOpen
  \bibfield  {author} {\bibinfo {author} {\bibfnamefont {J.-P.}\ \bibnamefont {Berenger}},\ }\bibfield  {title} {\bibinfo {title} {Perfectly matched layer for the fdtd solution of wave-structure interaction problems},\ }\href@noop {} {\bibfield  {journal} {\bibinfo  {journal} {IEEE Transactions on antennas and propagation}\ }\textbf {\bibinfo {volume} {44}},\ \bibinfo {pages} {110} (\bibinfo {year} {1996})}\BibitemShut {NoStop}%
\bibitem [{\citenamefont {Milton}\ \emph {et~al.}(2006)\citenamefont {Milton}, \citenamefont {Briane},\ and\ \citenamefont {Willis}}]{milton2006cloaking}%
  \BibitemOpen
  \bibfield  {author} {\bibinfo {author} {\bibfnamefont {G.~W.}\ \bibnamefont {Milton}}, \bibinfo {author} {\bibfnamefont {M.}~\bibnamefont {Briane}},\ and\ \bibinfo {author} {\bibfnamefont {J.~R.}\ \bibnamefont {Willis}},\ }\bibfield  {title} {\bibinfo {title} {On cloaking for elasticity and physical equations with a transformation invariant form},\ }\href@noop {} {\bibfield  {journal} {\bibinfo  {journal} {New Journal of Physics}\ }\textbf {\bibinfo {volume} {8}},\ \bibinfo {pages} {248} (\bibinfo {year} {2006})}\BibitemShut {NoStop}%
\end{thebibliography}%
%TC:endignore

\end{document}

\title{Supplemental Material:
Transformation twinning to create isospectral cavities}

\author{Simon~V~Lenz}
\affiliation{School of Biological Sciences, University of Bristol, BS8~1TQ, Bristol, UK}

\author{S\'ebastien~Guenneau}
\affiliation{Department of Mathematics, Imperial College London, SW7~2AZ, London, UK}
\affiliation{UMI 2004 Abraham de Moivre-CNRS, Imperial College London, SW7~2AZ, London, UK}

\author{Bruce~W~Drinkwater}
\affiliation{Department of Mechanical Engineering, University of Bristol, BS8~1TQ, Bristol, UK}

\author{Richard~V~Craster}
\affiliation{Department of Mathematics, Imperial College London, SW7~2AZ, London, UK}
\affiliation{UMI 2004 Abraham de Moivre-CNRS, Imperial College London, SW7~2AZ, London, UK}
\affiliation{Department of Mechanical Engineering, Imperial College London,  SW7~2AZ, London, UK}

\author{Marc~W~Holderied}
\affiliation{School of Biological Sciences, University of Bristol, BS8~1TQ, Bristol, UK}

\date{\today}

\begin{abstract}
In this supplementary material we provide the details for the derivation of material parameters characterizing the transformed media (carpet cloaks) coating the altered drum.
Given the practical difficulties in finding, or constructing, anisotropic media, but the ability to approximate them using effective medium theory and laminates of isotropic media we proceed to create effective media to illustrate the potential to build twinned cavities.
We identify the geometric and material parameters of  layered isotropic media approximating the material properties of transformed anisotropic media in the canonical case of carpet tents. Furthermore we provide numerical demonstration of twinning using this layered medium.
We highlight the meaning of parameters appearing in the transformed eigenvalue problem depending upon the physical setting, with particular attention paid to electromagnetic cavity twinning.
\end{abstract}

\maketitle

\section{Material properties of anisotropic media within twinned cavity}

\subsection{Data analysis software}

The boundary functions ($y_0$, $y_1$ \& $y_2$) for the transformations were defined in \python{} using the \emph{sympy} package \citep{sympy} and then compiled as a \clang{} library for use with \python{} and \comsol{}.
For all the finite element simulations, data was exported from \comsol{} to be analysed using \python{} with the \emph{numpy}, \emph{scipy} and \emph{matplotlib} packages \citep{numpy, scipy, matplotlib}.

\subsection{Further eigenfrequencies}

\begin{figure*}
  \includegraphics[width=\textwidth]{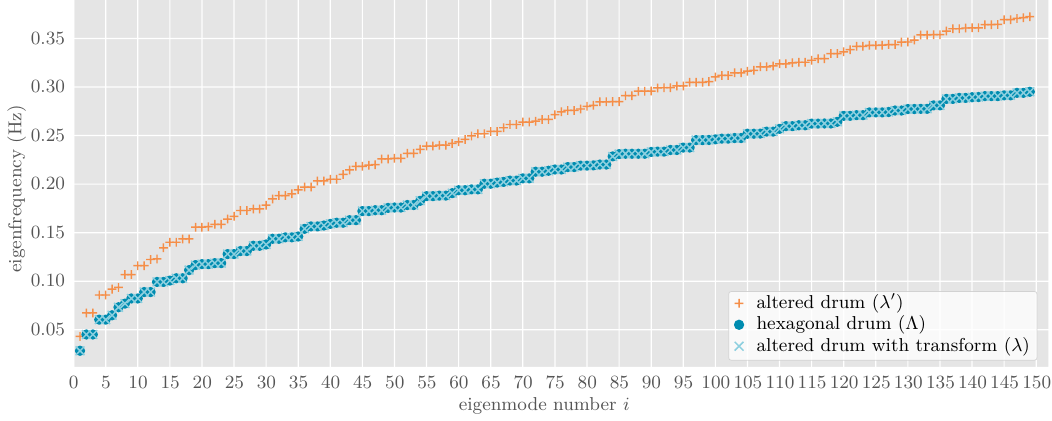}
  \caption{%
    Numerical comparison of the first 150 computed eigenfrequencies (some of which are degenerate) of the hexagonal drum versus the altered and the twinned drum assuming wave speed $c = 1\;ms^{-1}$, as $\sigma=\eta=1$ in the unaltered medium. %
    The eigenfrequencies of the altered drum are generally higher than the hexagonal drum. %
    With the transformation however, the eigenfrequencies of the altered drum match those of the hexagonal drum closely and achieve twinning. %
    The mean absolute difference between twinned and original drum is %
    $3.16e\text{-}7\pm3.68e\text{-}7$~Hz, %
    and $5.53e\text{-}2\pm 1.45e\text{-}2$~Hz between altered and original drum, %
    for a mesh with 2,500,000 elements (mean $\pm$ standard deviation).%
  }%
  \label{fig:freq_match_150}
\end{figure*}

Figure~\ref{fig:freq_match_150} shows the frequency range of $0-0.38$~Hz being twinned using the transformation approach for the meter-size hexagonal and altered drums in the Letter.
The mean absolute difference between the original and altered drums further increases as the spectra of the two drums further diverge at higher frequencies. Due to the linearity of the eigenvalue problems which we consider, this frequency range can be straightforwardly scaled up or down depending upon the size of the drums, and physical parameters.

\subsection{Polygonal cavities}
Figure \ref{fig:ap_anisotropy} shows a region sandwiched between the outer ($y_2$) and ground boundary ($y_0=0$) that is compressed into the region between outer ($y_2$) and inner ($y_1$) boundary.
We introduce a transformation mapping the region enclosed between two curves $(x,0)$ and $(x,y_2(x))$ to the one comprised between $(x,y_1(x))$ and $(x,y_2(x))$.
This corresponds to a compression of space from the $y_0$-$y_2$ region into the $y_1$-$y_2$ region.
In Figure \ref{fig:ap_anisotropy}, $(x,0)$ (ground boundary) is mapped on $(x,y_1(x))$ (inner boundary) and  $(x,y_2(x))$ (outer boundary) is fixed point-wise, using a mapping of the form

\begin{equation}
\left\{
\begin{array}{lr}
u=x\\
v=\alpha(x) y+ \beta(x)
\end{array}
\right.%
\label{eq:carpet}
\end{equation}

where
$\alpha (x)= ({y_2-y_1})/{y_2}$ and $\beta(x)= y_1$.

The inverse of this transformation is given by
\begin{eqnarray}\label{eq:inversetransfo}
  \left\{
  \begin{array}{lr}
    x=u\\
    y=({v-\beta(u)})/{\alpha (u)}
  \end{array}
  \right.
\end{eqnarray}

and the transformation \ref{eq:inversetransfo} has the following Jacobian matrix
\begin{eqnarray}
  \mathbf{J}_{\bf Xs}=\frac{\partial (x,y)}{\partial (u,v)}
  = \left(%
  \begin{array}{cc}
    \frac{\partial x}{\partial u} &  \frac{\partial x}{\partial v}\\
    \frac{\partial y}{\partial u} &  \frac{\partial y}{\partial v}\\
  \end{array}%
  \right) = \left(%
  \begin{array}{cc}
    1 &0\\
    g & \frac{1}{\alpha}\\
  \end{array}
  \right)
  \label{jacobian}
\end{eqnarray}

in which we have set

\begin{equation}
\begin{array}{lll}
g&:=&\displaystyle{\frac{\partial y }{\partial u} =\frac{1}{\alpha^2}
\left(-\alpha\frac{d\beta}{du}-(v-\beta)\frac{d\alpha}{du}\right)}
\\
&=&
\displaystyle{-\frac{1}{\alpha}\frac{dy_1}{du}-\frac{(v-y_1)}{\alpha^2}\frac{d}{du} \left( \frac{y_2-y_1}{y_2} \right)} \\
&=&\displaystyle{-\frac{(y_2-v)y_2}{(y_2-y_1)^2}\frac{dy_1}{du}+\frac{(y_1-v)y_1}{(y_2-y_1)^2}\frac{dy_2}{du}.}
\end{array}
\label{g}
\end{equation}

Hence we get the transformed parameter, $\sigma$, required in Eq. (3) of the Letter, given as

\begin{eqnarray}
  \label{eq:a}
  \sigma
  &=\mathbf{J}_{\bf sX}\mathbf{J}_{\bf sX}^{T}\det(\mathbf{J}_{\bf Xs}) \\
  &=
  \alpha
  \left(
  \begin{array}{cc}
  \frac{1}{\alpha} & 0\\
  - g & 1 \\
  \end{array}
  \right)
  \alpha
  \left(
  \begin{array}{cc}
  \frac{1}{\alpha} & -g\\
  0 & 1 \\
  \end{array}
  \right)
  \frac{1}{\alpha}
  \nonumber \\
  &=
  \left(
  \begin{array}{cc}
  \frac{1}{\alpha} & - g\\
  - g & (1+g^2) \alpha\\
  \end{array}
  \right)
  ,
\end{eqnarray}

and the transformed parameter, $\eta$, required in Eq. (3) of the Letter, as

\begin{equation}
  \label{eq:b}
  \eta
  =\; {\rm det}\mathbf{J}_{\bf Xs}=1/\alpha
\end{equation}

We can similarly design a transformation for all six boundaries of the twisted drum and turn this cavity into a hexagonal drum with the acoustics of the hexagonal drum.
To realise this practically this requires coating the walls of the twisted drum with metamaterials consisting of  layered isotropic media, as we shall see in the next section.

Before we move to the next section, we point out that we are not restricted to twinning of polygonal cavities, as we can map a curved boundary onto another curved boundary, instead of a curved boundary onto a flat boundary. In this way, we can twin a closed cavity of a given shape with a closed cavity of another shape, provided their boundaries can be represented by spline curves.

\subsection{Cavities with boundary defined by spline curves}
We now consider a region between the outer ($y_2$) and ground (curved) boundary ($y_0$), as shown in Figure \ref{fig:ap_anisotropy}, that is compressed into the region between outer ($y_2$) and inner ($y_1$) boundary.
We introduce a transformation that maps the region enclosed between two curves $(x,y_0(x))$ and $(x,y_2(x))$ to the one comprised between $(x,y_1(x))$ and $(x,y_2(x))$.
This corresponds to a compression of space from the $y_0$-$y_2$ region into the $y_1$-$y_2$ region.

More precisely, a curved ground boundary $(x,y_0)$ is mapped on $(x,y_1(x))$ (inner boundary) and  $(x,y_2(x))$ (outer boundary) is fixed point-wise, of the form \ref{eq:carpet}
where we now have
$\alpha (x)= ({y_2-y_1})/{(y_2-y_0)}$ and $\beta(x)= y_2(y_1-y_0)/{(y_2-y_0)}$. It is interesting to compare these expressions of $\alpha$ and $\beta$ with those in \ref{eq:carpet} when $y_0=0$.

The inverse of the transformation is given again by
\ref{eq:inversetransfo} which has the Jacobian matrix
\ref{jacobian}. However, we now have
\begin{equation}
    \begin{array}{lll}
g&:=&\displaystyle{\frac{\partial y }{\partial u} =\frac{1}{\alpha^2}
\left(-\alpha\frac{d\beta}{du}-(v-\beta)\frac{d\alpha}{du}\right)}
\nonumber\\
%&=&
%-\frac{1}{\alpha}\frac{dy_1}{du}-\frac{(v-y_1)}{\alpha^2}\frac{d}{du} \left( \frac{y_2-y_1}{y_2} \right) \nonumber\\
%&=&
%-\frac{y_2}{y_2-y_1}\frac{dy_1}{du} \nonumber \\
%&+&\frac{(y_1-v)y_2^2}{(y_2-y_1)^2}\left ( -\frac{1}{y_2}\frac{dy_1}{du} + \frac{y_1}{y_2^2} \frac{dy_2}{du} \right) \nonumber\\
%&=&-\frac{dy_1}{du} \left( \frac{y_2}{y_2-y_1}+\frac{(y_1-v)y_2^2}{y_2(y_2-y_1)^2}\right) \nonumber \\
%&+&\frac{(y_1-v)(y_1)}{(y_2-y_1)^2}\frac{dy_2}{du} \nonumber \\
&=&\displaystyle{-\frac{(y_2-v)(y_2-y_0)}{(y_2-y_1)^2}\frac{dy_1}{du}\nonumber+\frac{(y_1-v)(y_1-y_0)}{(y_2-y_1)^2}\frac{dy_2}{du}.}
\end{array}
\label{gbis}
\end{equation}

This expression should be compared with \ref{g}.
We deduce the transformed parameters, $\sigma$ and $\eta$, given again as \ref{eq:a} and \ref{eq:b}, respectively.

We can thus twin cavities with complex shapes. However, the more complex their boundaries, the more complex are the metamaterials designed to approximate the required transformed media.
Therefore, for practical implementation, it is easier to consider specific geometries allowing for simpler material parameters.

\subsection{Cavities with boundary defined by carpet tents}
\begin{figure*}
    \centering
    \includegraphics[width=\textwidth]{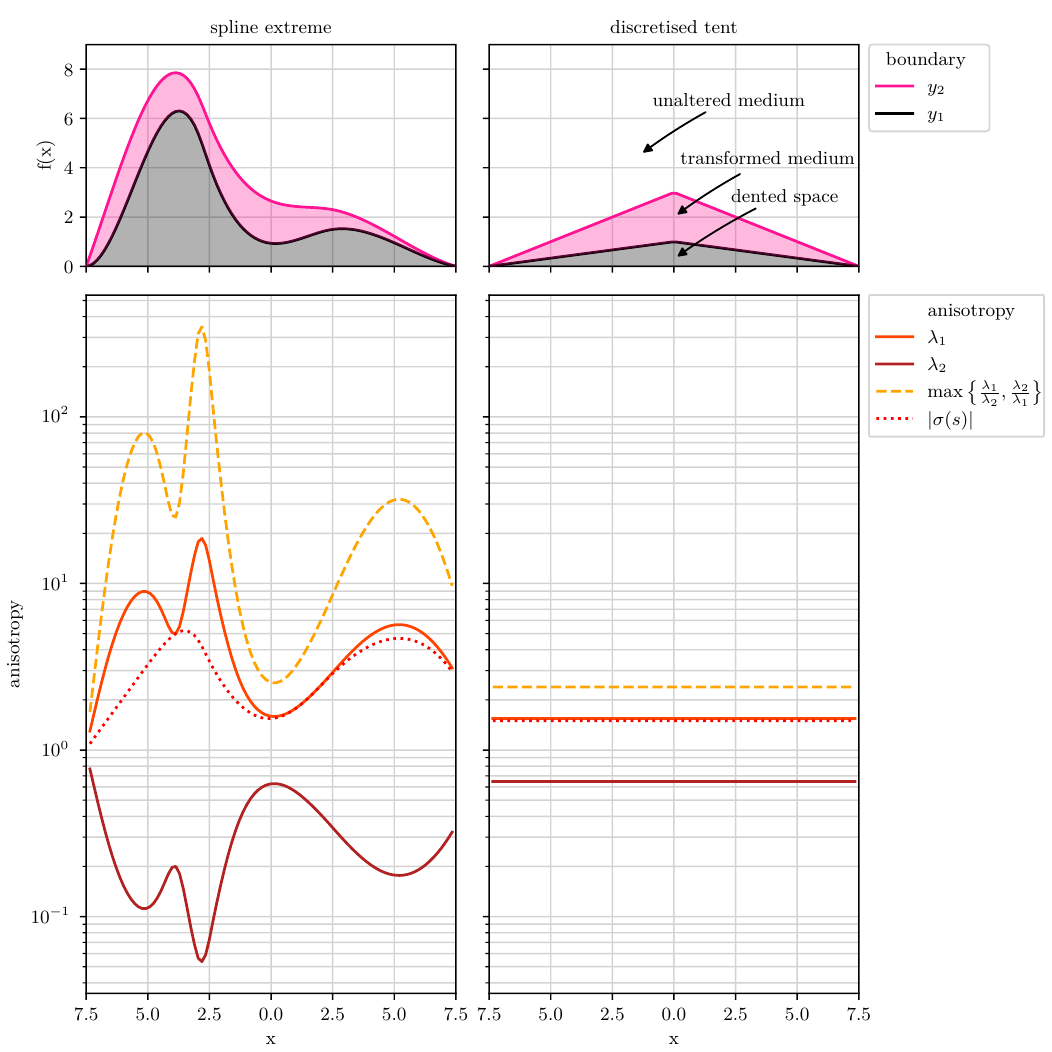}
    \caption{
    Linear geometric transforms, \autoref{eq:carpet}, and associated eigenvalues of coefficient $\sigma$ in \autoref{eq:a}:
    \textbf{Top:} geometric representation of cavity boundaries.
    As part of the geometric transform, the region between outer ($y_2$) and ground boundary ($y_0=0$) is compressed into the region between outer ($y_2$) and inner ($y_1$) boundary.
    \textbf{Bottom:} eigenvalues of matrix $\sigma$, and their ratio, for the respective geometric transforms.
    We note that no eigenvalues vanish and thus $\sigma$ is positive definite and bounded.
    These eigenvalues are a representative measure for the anisotropy needed in $\sigma$ to achieve an isospectral transformation.
    Note that the required anisotropy for the extreme spline case might be achievable in practice in electromagnetism via effective medium theory with alternating layers of e.g. fused silica and BSTO-oxide composite ceramics that have a dielectric constant ranging  respectively  from 3.5 to 4 and 100 to 640 in the microwave (GHz) regime \cite{o2002photonic}. Further note the transverse magnetic (TM) polarization prevents magnetism to enter the transformed eigenvalue equation
    (see Eq. \ref{eq:prop4} and Table \ref{tab:physics}) if some star-shape cavity is assumed, in which case an infinite conducting boundary corresponds to Neumann datum (Dirichlet datum holds for the transverse electric polarization).}
    \label{fig:ap_anisotropy}
\end{figure*}

Of particular interest are cavities with boundaries defined by carpet tents as shown in the right panel of Figures \ref{fig:ap_anisotropy} and \ref{fig:ap_discretised_modes}. In that star-shape geometry, we note that $\eta
  =\; {\rm det}\mathbf{J}_{\bf Xs}=1/\alpha$ in Eq. \ref{eq:b} is constant, and thus the transformed eigenvalue problem Eq. (3) in the Letter
 %TC:ignore
\begin{equation}
-\nabla_{\textbf{s}}\cdot ( \sigma(\textbf{s}) \nabla_{\textbf{s}} \phi(\textbf{s})) = \eta \lambda \phi(\textbf{s}) \;
\label{eq:prop2}
\end{equation}
%TC:endignore
where $\sigma$ is given by Eq. \ref{eq:a} can be recast as
\begin{equation}
-\nabla_{\textbf{s}}\cdot \left( \frac{\sigma(\textbf{s})}{\eta} \nabla_{\textbf{s}} \phi(\textbf{s}) \right) =\lambda \phi(\textbf{s}) \;
\label{eq:prop4}
\end{equation}
%in the transformed domain $\pi$
where $\nabla_{\textbf{s}}$ is the gradient operator in $u,v$ coordinates (indeed, eigenfunctions $\phi(\textbf{s}(\textbf{X}))= \Phi(\textbf{X})$ are defined up to a multiplication constant so it is legitimate to multiply Eq. \ref{eq:prop2} throughout by the constant $\eta^{-1}$).

This is particularly useful in the case of a transverse magnetic (TM) cavity, as this prevents magnetism to enter the transformed equation, see Table III.

We now proceed to approximate the transformed media by layered metamaterials for cavities with a star-shape geometry.

\begin{table}
    \centering
    \begin{tabular}{ccccc}
  x (m) & $y_1$ (m) & $y_2$ (m) & $\sigma$                              & $\eta$ \\
-7.5000 & 0.0000    & 0.0000    & NaN & 1.9231 \\[.5cm]
-6.5000 & 1.1200    & 2.9200    & $\left(\begin{array}{cc} 1.6222 & 3.2792               \\ 3.2792   & 7.2453                            \end{array}\right)$ & 1.6222 \\[.5cm]
-3.0000 & 5.5000    & 7.0000    & $\left(\begin{array}{cc} 4.6667 & -20.7191             \\ -20.7191 & 92.2033                           \end{array}\right)$ & 4.6667 \\[.5cm]
-2.5000 & 4.1000    & 5.8000    & $\left(\begin{array}{cc} 3.4118 & -16.6975             \\ -16.6975 & 82.0126                           \end{array}\right)$ & 3.4118 \\[.5cm]
2.5000  & 1.5000    & 2.3000    & $\left(\begin{array}{cc} 2.8750 & 0.8874               \\ 0.8874   & 0.6217                            \end{array}\right)$ & 2.8750 \\[.5cm]
6.5000  & 0.3000    & 0.4000    & $\left(\begin{array}{cc} 4.0000 & -0.8444              \\ -0.8444  & 0.4282                            \end{array}\right)$ & 4.0000 \\[.5cm]
7.5000  & 0.0000    & 0.0000    & NaN & 3.0000 \\[.5cm]

    \end{tabular}
    \caption{Spline curve definition. Note that $y_0 = 0$ and $c = 1\;ms^{-1}$ in the unaltered medium (as a result of $\sigma = \eta = 1$). Values were evaluated at the cubic spline nodes. The supplemental csv file provides 100 datapoints of the splines, matrix $\sigma$ and scalar $\eta$. In the starting and end point of the carpet splines, the transformed parameter $\sigma$ is undefined (NaN), as $y_0 = y_1 = y_2 = 0$.}
    \label{tab:ap_spline_vals}
\end{table}

\section{Design of anisotropic media via effective medium theory}
A canonical example of periodic structure that can be used to approximate the properties of the transformed medium characterized by parameters $\sigma$ and $\eta$ (see. Eq. \ref{eq:a}-\ref{eq:b}) is an alternation of isotropic homogeneous layers.
However, one should note that, in general, these layers should have different orientations in different parts of the transformed medium, as $\sigma$ and $\eta$ are spatially varying (see Fig. \ref{fig:ap_anisotropy}) and thus we consider here polycrystals with piecewise constant and scalar parameters $\sigma_i$ and $\eta_i$.
Analytical formulae can then be employed to compute the effective parameters of such effective anisotropic media \cite{Bensoussan1978,Milton2002}:

\begin{figure*}
    \centering
    \includegraphics[width=\textwidth]{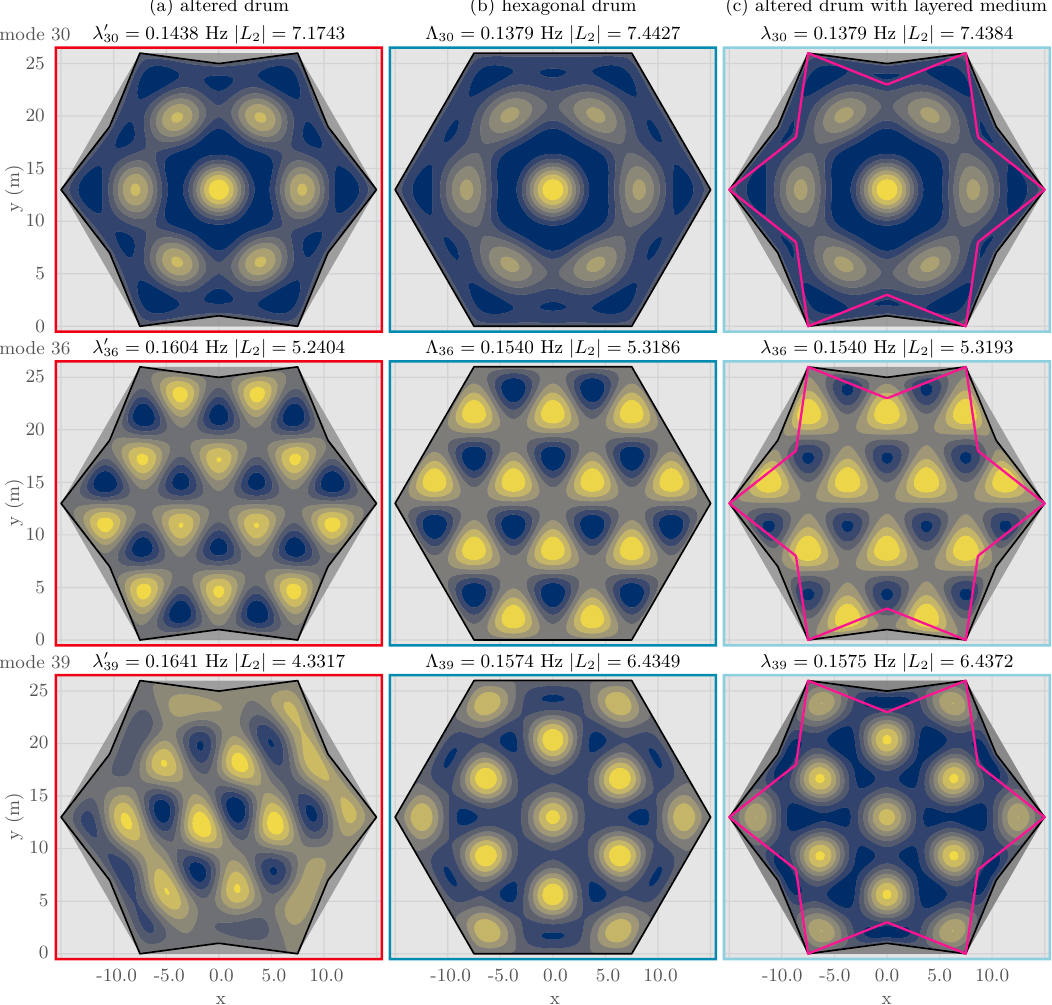}
    \caption{Selection of modes for a drum with layered metamaterial (note that in the unaltered medium, $c=1\;ms^{-1}$ as a result of $\sigma=\eta=1$).
    The eigenmodes and eigenfrequencies of the altered drum with a tent-shaped inclusion (a) are different to the hexagonal drum (b).
    Introducing a layered medium within a boundary layer (between the black and pink line) of the altered drum (c) restores the modes and eigenfrequencies of the hexagonal drum (b).
    For each mode and drum, the eigenfrequencies and $L_2$ norm are provided above each panel.
    The metamaterial consists of 20 parallel layers in each boundary section (240 total) and there were 148,302 elements in the mesh.}%
    \label{fig:ap_discretised_modes}
\end{figure*}

\begin{equation}
\Sigma =
\left(
\begin{array}{ll}
\Sigma_{xx}  & \Sigma_{xy} \\
\Sigma_{yx} & \Sigma_{yy}
\end{array}
\right)
=
{\bf R}(\theta)\left(
\begin{array}{ll}
\langle \sigma_i^{-1} \rangle ^{-1}  & 0 \\
0 & \langle \sigma_i \rangle
\end{array}
\right){\bf R}(-\theta)
\label{eq:param_hom1Da}
\end{equation}
where ${\bf R}(\theta)$ is the rotation matrix through an angle $\theta$, and
\begin{equation}
\eta = \langle \eta_i \rangle
\label{eq:param_hom1Db}
\end{equation}
where $\langle . \rangle$ denotes the mean over a period in the direction orthogonal to the interface between layers.

Our algorithm to approximate the required artificial anisotropic parameters is well-known, so for simplicity we consider here the case of tent carpet cloaks.
In this configuration the tuning parameters are the different homogeneous isotropic media defined by $(a_i,b_i)$ in the periodic cell.
Possible additional tuning parameters are the size $L_0$ and $L_1$ of each layer.
%In the following numerical results we choose to set $L_0 = L_1$.
Assuming for simplicity that $\theta=0$ (so $\Sigma$ is a diagonal matrix), we derive the numerical value of $\sigma_1$ and $\sigma_2$ by solving the following system \cite{Petiteau2014}:

\begin{equation}
\begin{array}{ll}
\Sigma_{xx} &= \displaystyle{\langle \sigma_i^{-1} \rangle ^{-1} = \frac{\sigma_0 \sigma_1}{\sigma_1 L_0 + \sigma_0 L_1}} \\
\Sigma_{yy} &= \langle \sigma_i \rangle = L_0 \sigma_0 + L_1 \sigma_1 \\
\eta &= \langle \eta_i \rangle = L_0 \eta_0 + L_1 \eta_1
\end{array}
\label{homlayered}
\end{equation}

$L_0$ and $L_1$ being the thickness of the two layers that compose the microstructure in a given part of the polycrystal.

As a side note, effective medium formulae \ref{homlayered} do not include a dynamic correction which can be safely neglected at low frequencies, but would play a role for moderate frequencies.
Higher-order homogenization formulae can be implemented and the corresponding effective medium will be described by a frequency dependent $\Sigma$ (i.e.
dispersive medium) \cite{Milton2002}, which makes the counterpart of eigenvalue problem \ref{eq:prop4} for the homogenized layered medium, non-linear.
Whilst we are aware of this limitation of the approximation of transformed media by layered media, we have numerically checked that this does not affect the cavity twinning for the first few tens of eigenvalues, as reported in the Letter and Figure S4.

Of course, one could design more complex structured media approximating the required parameters in \ref{eq:a} and \ref{eq:b}. This can be done for instance considering some inverse homogenization approach such as in \cite{pomot2020acoustic} that applies a genetic algorithm to design circular cylindrical cloaks. This approach should work also for our twinning cavity problem, albeit it being an eigenvalue problem, rather than a scattering problem.
Let us now discuss this aspect of homogenization applied to cavity problems.

\begin{table}[]
    \centering
    \begin{equation*}
        \sigma_{tent} =
        \left(
        \begin{array}{ll}
        1.50  & 0.20 \\
        0.20 & 0.693
        \end{array}
        \right)\qquad
        \eta_{tent} = 1.50
    \end{equation*}
    \begin{tabular}{cccccccc}
        $\sigma_0$ & $\sigma_1$ & $\frac{L_0}{L_1}$ & layer orientation $\theta$ ($^\circ$) \\ \hline
        2.93 & 0.393 & 0.832 & 13.18 \\
    \end{tabular}
    \caption{Properties of the layered medium. \textbf{Top:} transformed parameters $\sigma$ and $\eta$ for the tent-shaped transformation. \textbf{Bottom:} design parameters for the layered medium to approximate the transformed parameters. $\sigma_0$ and $\sigma_1$ describe the parameter $\sigma$ for the two alternating layers, respectively. The thickness of the layers is provided as a dimensionless ratio. Within the medium the layers are oriented at the angle $\theta$ against the horizontal.}
    \label{tab:ap_metamat_vals}
\end{table}

\subsection{Homogenized Maxwell's equations}
In a region with no charges and no currents, the homogenized Maxwell's equations take the form:
\begin{equation}
{\displaystyle {\begin{aligned}\nabla \cdot (\varepsilon_0\underline{\underline{\varepsilon}}_{\rm eff}\mathbf {E}) &=0,&\nabla \times \mathbf {E} &=-{\frac {\partial \mathbf {B} }{\partial t}},\\\nabla \cdot \mathbf {B} &=0,&\nabla \times (\mu_0^{-1} \underline{\underline{\mu}}^{-1}_{\rm eff}\mathbf {B}) &=\varepsilon _{0}\underline{\underline{\varepsilon}}_{\rm eff}{\frac {\partial \mathbf {E} }{\partial t}}.
\end{aligned}}}
\label{maxfull}
\end{equation}
with $\underline{\underline{\varepsilon}}_{\rm eff}$ and $\underline{\underline{\mu}}_{\rm eff}$ the homogenized rank-2 tensors of relative permittivity and permeability, respectively (see \cite{guenneau2000homogenization,kristensson2003homogenization,guenneau2007homogenization} for associated local cell problems), $\mathbf {E}$ the electric field, $\mathbf {B}$ the magnetic field and $\varepsilon_{0}$ (resp. $\mu _{0}$) is the vacuum permittivity (resp. permeability).

At the same time, the electric displacement field $\mathbf {D}$ is related to the electric field $\mathbf {E}$, and the magnetic field $\mathbf {B}$ is related to the magnetization field $\mathbf {H}$, by the following relations:
\begin{equation}{\displaystyle {\begin{aligned}
\mathbf {D} &=\ \varepsilon _{0}\mathbf {E} +\mathbf {P} \ =\ \varepsilon _{0}({\bf I}+\underline{\underline{\chi}}_{\text{e}})\mathbf {E} \ =\ \varepsilon _{0}\underline{\underline{\varepsilon}}_{\rm eff}\mathbf {E} \\
\mathbf {B} &=\ \mu _{0}\left(\mathbf {H} +\mathbf {M} \right)\ =\ \mu _{0}\left({\bf I}+\underline{\underline{\chi}}_{\text{m}}\right)\mathbf {H} \ =\ \mu_0 \underline{\underline{\mu}}_{\rm eff}\mathbf {H} \end{aligned}}}
\label{polarization}
\end{equation}
where $\mathbf {P}$ (resp. $\mathbf {M}$) is the polarization (resp. magnetization) field and $\underline{\underline{\chi}}_{\text{e}}$ (resp. $\underline{\underline{\chi}}_{\text{m}}$) is the rank-2 tensor of electric (resp. magnetic) susceptibility. Moreover, $\underline{\underline{\varepsilon}}_{\rm eff}$ (resp. $\underline{\underline{\mu}}_{\rm eff}$) is the homogenized rank-2 tensor of relative permittivity (resp. permeability). We note that $\underline{\underline{\chi}}_{\text{e}}$ and $\underline{\underline{\chi}}_{\text{m}}$ appear as correctors in the definition of homogenized permittivity $\underline{\underline{\varepsilon}}_{\rm eff}$ and permeability $\underline{\underline{\mu}}_{\rm eff}$ in \cite{guenneau2000homogenization,kristensson2003homogenization,guenneau2007homogenization} and they depend on local cell problems.  

The electric (resp. magnetic) susceptibility of most metamaterials is not a scalar quantity. Indeed, the electric (resp. magnetic) response $\mathbf {P}$ (resp. $\mathbf {M}$) is dependent upon the orientation of the sample and can occur in directions other than that of the applied electric (resp. magnetization) field $\mathbf {E}$ (resp. $\mathbf {H}$). In these cases, electric and magnetic susceptibilities $\underline{\underline{\chi}}_{\text{e}}$ and $\underline{\underline{\chi}}_{\text{m}}$ are defined as
\begin{equation}
P_{i}=E_{j}\chi_{\text{e}}^{ij} \; , \; M_{i}=H_{j}\chi_{\text{m}}^{ij}
\end{equation}
where $i$ and $j$ refer to the directions (e.g. $x$ and $y$ Cartesian coordinates) of the applied field and polarization (resp. magnetization), respectively. These rank-2 tensors describe the components of polarization (resp. magnetization) in the $i$th direction from the external electric (resp. magnetization) field applied in the $j$th direction.

Assuming a time-harmonic dependence $\exp(-i\omega t)$, with $\omega$ the angular frequency in \ref{maxfull}, it is easily seen that eigenfrequencies and associated eigenmodes of a three-dimensional closed electromagnetic cavity filled with the homogenized medium are solutions of
\begin{equation}
\left\{
\begin{array}{lll}
 \nabla\times {(\underline{\underline{\mu}}_{\rm eff})}^{-1}\nabla\times {\bf E}(x,y,z) = \varepsilon_{0}\mu_{0}{\varepsilon}_{\rm eff}\omega^2 {\bf E}(x,y,z)
 & \\
\nabla\times {(\underline{\underline{\varepsilon}}_{\rm eff})}^{-1} \nabla\times {\bf H}(x,y,z) = \varepsilon_{0}\mu_{0}{\mu}_{\rm eff}\omega^2 {\bf H}(x,y,z)
&
\end{array}
\right.
\label{maxcavity1-3D}
\end{equation}
One need only numerically solve one of these two homogenized equations, subject to infinite conducting boundary conditions. It is well known the spectrum consists of a countable set of real positive eigenfrencies \cite{bossavit1990solving,zolla2005foundations}. If the medium within the cavity is non-magnetic, it is customary to solve the eigenvalue equation with the magnetic field as the unknown. Indeed, the magnetic field is divergence free (as can be seen applying the divergence to the second equation). Making use of edge-based finite elements (also known as Nedelec elements \cite{monk2003finite}) then avoids the appearance of spurious modes when computing eigenfrequencies and associated eigenmodes in closed electromagnetic cavities \cite{bossavit1990solving}.
If the medium within the cavity is described by effective permittivity and permeability tensors which are both different from the identity tensor, then one needs to make use of properties of weigthed anisotropic Hilbert spaces to check good properties are preserved, and there is no spectral pollution. We will not comment any further on the twinning of three-dimensional electromagnetic closed cavities as this falls beyond the scope of the Letter. 

%However, for our purpose of cavity twinning, ${\mu}_{\rm eff}$ is generally anisotropic and heterogeneous. So-called finite edge elements can nonetheless be employed to accurately solve such eigenvalue problems, see e.g. \cite{zolla2005foundations}.    

For a two-dimensional electromagnetic cavity in the $(xy)$-plane filled with the effective medium, \ref{maxcavity1-3D} reduce to the anisotropic Helmholtz equations
\begin{equation}
\left\{
\begin{array}{lll}
 -\nabla\cdot {(\underline{\underline{\mu}}_{\rm eff}^{2D})}^{-1} \nabla E_z(x,y) = \varepsilon_{0}\mu_{0}({\varepsilon}_{\rm eff}^{33})\omega^2 E_z(x,y)
 & \hbox{(TE)} \\
-\nabla\cdot {(\underline{\underline{\varepsilon}}_{\rm eff}^{2D})}^{-1} \nabla H_z(x,y) = \varepsilon_{0}\mu_{0}({\mu}_{\rm eff}^{33})\omega^2 H_z(x,y)
& \hbox{(TM)}
\end{array}
\right.
\label{maxcavity1}
\end{equation}
where $E_z$ and $H_z$ are the out-of-plane components of the electric and magnetic field, in transverse electric (TE) and transverse magnetic (TM) polarizations respectively, and $\underline{\underline{\varepsilon}}_{\rm eff}^{2D}$ and ${\varepsilon}_{\rm eff}^{33}$ (resp. $\underline{\underline{\mu}}_{\rm eff}^{2D}$ and ${\mu}_{\rm eff}^{33}$) are the block diagonal parts of the tensors of homogenized permittivity (resp. permeability) defined as
\begin{equation}
\underline{\underline{\varepsilon}}_{\rm eff} =
\left(
\begin{array}{ll}
\underline{\underline{\varepsilon}}_{\rm eff}^{2D}  & {\bf 0} \\
{\bf 0}^T & {\varepsilon}_{\rm eff}^{33}
\end{array}
\right)
\; , \;
\underline{\underline{\mu}}_{\rm eff}^{2D} =
\left(
\begin{array}{ll}
\underline{\underline{\mu}}_{\rm eff}^{2D}  & {\bf 0} \\
{\bf 0}^T & {\mu}_{\rm eff}^{33}
\end{array}
\right)
\label{maxcavity2}
\end{equation}
with ${\bf 0}$ the null 2-component column vector and ${\bf 0}^T$ its transpose.
One can easily recast \ref{maxcavity1} in terms of the electric and magnetic susceptibities using \ref{polarization} and \ref{maxcavity2}. The form of permittivity and permeability tensors in \ref{maxcavity2} is known as z-anisotropy in the context of crystal fibres, and this structure is preserved in the fully vectorial case \cite{zolla2005foundations}. This can be useful for twinning waguides.

Importantly, $\underline{\underline{\varepsilon}}_{\rm eff}^{2D}$ and $\underline{\underline{\mu}}_{\rm eff}^{2D}$ in \ref{maxcavity2} are straightforwardly identified to the inverse of $\Sigma$ in \ref{eq:param_hom1Da} and ${\varepsilon}_{\rm eff}^{33}$ and ${\mu}_{\rm eff}^{33}$ are identified to $<\eta>$ in \ref{eq:param_hom1Db}. Therefore, one can use the effective medium formulae \ref{homlayered} to generate the required artificial anisotropic medium for the two-dimensional electromagnetic cavity with isotropic layers. Of course, in order to achieve layers with some magnetic activity, one needs to either resort to high-contrast dielectric layers, or to layers themselves consisting of split ring resonators or the like. An other option is to use some reduced parameters in TM polarization, what amounts to dividing the second equation in \ref{maxcavity1} by ${\mu}_{\rm eff}^{33}$ and considering an effective permivitty $\underline{\underline{\varepsilon}}_{\rm eff}^{2D}$, similarly to was done in \ref{eq:prop2} to get \ref{eq:prop4}. In this way, dielectric layers can be used to achieve the required anisotropy in TM polarization (for an infinite conducting cavity with Neumann datum).

%In general, a material cannot polarize instantaneously in response to an applied field, and so the more general formulation as a function of time is
%\begin{equation}
%\mathbf {P} (t)=\varepsilon _{0}\int _{-\infty }^{t}\chi %_{\text{e}}(t-t')\mathbf {E} (t')\,\mathrm {d} t'
%\end{equation}
%That is, the polarization is a convolution of the electric field at previous times with time-dependent susceptibility given by $\chi _{\text{e}}(\Delta t)$.

\section{Homogenization theory for eigenvalue problems}
There is a vast amount of literature on effective medium theory \cite{Milton2002}, and it has been successfully applied to many wave physics problems ranging from acoustics, to elastodynamics and electromagnetics.
However, a word of caution is in order, since we are concerned here with spectral problems, which are markedly different from scattering problems.
Usually, researchers apply homogenization theory at a fixed wave frequency, in which case it is well known homogenization breaks down when the wavelength is comparable to the size of the elementary cell/ structural elements.
Provided homogenization is applicable (theoretically heterogeneities within the medium should be much smaller than the wave wavelength, or in other words the wave frequency should be nearly zero) the wavefields for the ideal medium and the structured medium that approximate the former, should be almost indistinguishable.

Mathematically, for a fixed frequency, when the elementary cell becomes smaller and smaller, the $L_2$ norm of the wavefields on every compact set in a bounded or unbounded domain should become closer and closer, see e.g.
\cite{guenneau2000homogenization,kristensson2003homogenization} for the case of electromagnetic waves.

However, we deal here with eigenvalue problems, and twinning cavities amounts to control over both the eigenvalues and associated eigenfields.
This branch of homogenization theory is concerned with convergence of spectra.
Clearly, homogenization should work for the lower part of the spectrum (There is nonetheless a way to break free of the constraint of a long wavelength compared to the structural element size, known as high frequency homogenization \cite{craster2010high}, but this left for further work.), and we shall thus aim at twinning the first few eigenvalues and associated eigenfields of cavities different shapes.

The mathematical framework can be found in the classical textbook \cite{Bensoussan1978}, so we just briefly recount essential results undepinning our twinning cavity theory.
It is customary to denote by $0<\delta\ll 1$ a small parameter that corresponds to the ratio betwen the size of a periodic elementary cell and that of the cavity.
When $\delta$ tends to zero, the cavity is then filled with a very large number of very small cells and it can be considered as filled with a homogeneous medium.
The material parameters of the heterogeneous medium are $\sigma_\delta$ and $\eta_\delta$.
These parameters are spatially varying (for instance in a stepwise fashion) and they should should be positive and bounded.
Let us consider eigenvalues $\Lambda_\delta$ and associated eigenvectors $\Phi_\delta$ of the heterogeneous Laplacian in a closed cavity:
\begin{equation}\label{eq:Peb}
\left\{
\begin{array}{lll}
 -\nabla\cdot \sigma_\delta(\textbf{X}) \nabla \Phi_\delta(\textbf{X}) = \eta_\delta(\textbf{X}) \Lambda_\delta \Phi_\delta(\textbf{X})
 & \hbox{, in
$\Pi$,} \\
{\Phi}_\delta = 0 \mbox{ or }{{\bf n}\cdot [\sigma_\delta\nabla\Phi}_\delta] = 0
& \hbox{, on
$\partial\Pi$.}
\end{array}
\right.
\end{equation}

where the boundary condition (Dirichlet or Neumann datum) depends upon the the physical setup (for the drum problem, this would be Dirichlet, and in electromagnetics, an infinite conducting cavity would assume the latter, Neumann, datum in transverse magnetic polarization).

It is well-known \cite{Bensoussan1978} that the operator associated with \ref{eq:Peb} has a compact resolvent.

We denote by ${\cal S}_{\delta}:= \{\Lambda_\delta^{(j)};
j=1,\dots \}$ the spectrum of the operator in \ref{eq:Peb}, which is a discrete countable set of eigenvalues tending to infinity (we note in passing this is no longer the case for an open cavity.)

Passing to the limit when $\delta$ tends to $0$, one obtains the limit eigenvalue problem \cite{Bensoussan1978}
\begin{equation}
\left\{
\begin{array}{lll}
 -\nabla\cdot \Sigma_0(\textbf{X}) \nabla \Phi_0(\textbf{X}) = \eta_0(\textbf{X}) \Lambda_0 \Phi_0(\textbf{X})
 & \hbox{, in
$\Pi$,} \\
{\Phi_0}= 0 \mbox{ or }{{\bf n}\cdot [\Sigma_0\nabla\Phi}_0] = 0 &
\hbox{, on
$\partial\Pi$,}
\end{array}
\right.
\label{eq:limitproblem}
\end{equation}
where $\Sigma_0$ and $\eta_0$ are given by $\rm{Diag}(\langle \sigma_i^{-1}\rangle^{-1},\langle\sigma_i\rangle)$ and $\langle \eta_i\rangle$, respectively, in the layered case. $\Sigma_0$ is thus a matrix, whereas $\eta_0$ remains scalar. The general expressions for $\Sigma_0$ and $\eta_0$ are given in \cite{Bensoussan1978}, and require solving a local problem on the elementary cell.
The resolvent of the operator associated with \ref{eq:limitproblem} is also compact, so the limit spectrum
${\cal S}_{0}:= \{\Lambda_0^{(j)};
j=1,\dots \}$
is made up of a countable set of discrete eigenvalues.
One can then write that each eigenvalue $\Lambda_\delta$ in ${\cal S}_{\delta}$ converges pointwise to an eigenvalue $\Lambda_0$ in ${\cal S}_{0}$.

The limit of the sequence of spectra $S_\delta$ turns out to satisfy \cite{Allaire1998}
\begin{equation}
{\cal S}_{0}\subset\lim_{\delta\to 0}{\cal S}_{\delta}
\label{eq:sigma0}
\end{equation}
where ${\cal S}_{0}$ is the homogenized part of the limit spectrum (associated with low frequency eigenmodes).
Likewise there is also convergence of sequences of eigenfields $\Phi_\delta$ to $\Phi_0$, and this is in L2 norm over the domain $\Pi$.

One cannot write an equality in \ref{eq:sigma0}.
For instance, when the domain $\Pi$ is not filled with an integer number of cells (as is the case for the altered drum with layered medium in Fig.
\ref{fig:ap_discretised_modes}), there might be some additional part to the homogenized spectrum associated with boundary layer effects.
This can be viewed as some spectral pollution induced by eigenvalues associated with eigenmodes concentrating near the boundary of the cavity (in other words, this spoils the twinning effect with unwanted eigenvalues in the spectrum of the altered drum with layered medium).
And besides from that, at higher frequencies, the limit spectrum $\lim_{\delta\to 0}{\cal S}_{\delta}$ is made up of bands see \cite{Allaire1998} (and thus this would again add some spectral pollution to the twinning of the hexagonal drum in Fig.
\ref{fig:ap_discretised_modes} with some unwanted bands in the higher frequency part of the spectrum of the altered drum with layered medium).

Therefore, our homogenization approach for cavity twinning only applies to the lower part of the spectrum.
However, high frequency homogenization \cite{craster2010high} might be a way to twin higher eigenfrequencies and associated eigenfields, but this is beyond the scope of this work.

\section{Numerical proof-of-concept cavity twinning}

One issue with the effective medium approach in the previous section is the number of layers required within each carpet cloak of the altered drum with layered medium in Fig.
\ref{fig:ap_discretised_modes}.
The number of layers, and their thickness underpins the approximation for a given number of eigenvalues (possibly of multiplicity greater than 1) and associated eigenfields.
From homogenization theory, it seems clear that the higher the number of eigenvalues one wishes to twin between the altered drum with layered medium and the reference hexagonal cavity in Fig.
\ref{fig:ap_discretised_modes}, the finer the layering within the carpet cloaks of the altered drum.

However, it is hard to provide a definite answer for the number of layers required for a given number of eigenvalues to twin in Fig.
\ref{fig:ap_discretised_modes}.
Indeed, previous works have focussed on cloaking efficiency in scattering problems, see e.g.
\cite{Petiteau2014} for a numerical analysis suggesting one might need as many as 10,000 layers to achieve nearly perfect cloaking with a cylindrical cloak consisting of concentric isotropic layers at a fixed frequency.
Here, we deal with the more challenging problem of approximating both eigenfrequencies and eigenfields of a reference cavity with an altered cavity dressed with metamaterials, so it is legitimate to wonder if one can achieve twinning with layered media available in practice.

\begin{figure}
    \centering
    \includegraphics[width=\columnwidth]{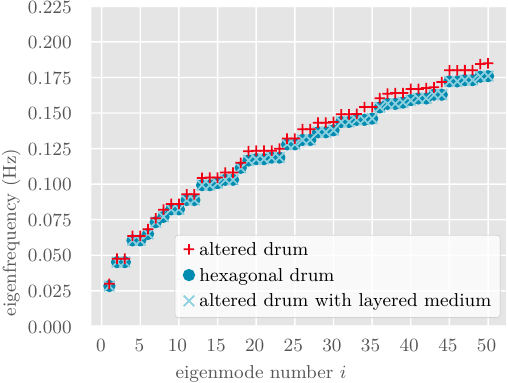}
    \caption{Eigenfrequencies of the drums with and without layered metamaterial (note that in the unaltered medium, $c=1\;ms^{-1}$ as $\sigma = \eta = 1$). The mean difference ($\pm$ standard deviation) between altered drum and original drum is $5.66e\text{-}03 \pm 1.913e\text{-}03$~Hz, while for the altered drum coated with layered media it is $5.09e\text{-}05 \pm 3.05e\text{-}05$~Hz. There were 20 layers in each separate boundary layer (240 layers total) and 148,302 elements in the mesh.}%
    \label{fig:ap_discretised_frequencies}
\end{figure}

We provide a positive answer for the first 50 eigenvalues and associated eigenfields for the case of a dented hexagonal cavity dressed with layered media consisting of 20 layers per section, see Fig.
\ref{fig:ap_discretised_modes} and Fig.
\ref{fig:ap_discretised_frequencies}.
In Table \ref{tab:ap_metamat_vals}, one can see that the numerical values for material parameters $\sigma_0$, $\sigma_1$ and $\eta$ are within the range of available materials in acoustics.
Indeed, $\sigma_0/\sigma_1\sim 7.45$ corresponds to a ratio of densities between layers, and one can achieve this ratio for instance with an alternation of coal and metal such as steel or iron.
One can also achieve a larger ratio with an alternation of cork and metal.
Regarding the parameter $\eta$, we chose for simplicity some layers with same bulk modulus as in Table \ref{tab:ap_metamat_vals}, but this can be fine tuned for the case of an alternation of coal/cork and metal in a way similar to what was done for acoustic carpet cloaks in \cite{popa2011experimental}.
In fact, the key point is to achieve the required anisotropy, and thus of foremost importance is the ratio of densities.
Likewise, to twin electromagnetic cavities one would design the carpet cloaks with specific dielectric layers, and here again a ratio of $\sigma_0/\sigma_1\sim 7.45$ is easily achievable since most dielectrics have a relative permittivity between between $1$ and $13$ (however, $\eta$ needs to be $1$ as it corresponds to relative permeability, and so the transverse magnetic TM case should be favored, see Table \ref{tab:physics}).
Electromagnetic carpet cloaks dressing the altered cavity would be designed for instance as in \cite{ergin2010three}, bearing in mind that perfect twinning can only be achieved for one light polarization (depending upon the light polarization, one would need to alternate layers with markedly different permeability, which is hard to achieve in practice, and so it seems more plausible to achieve twinning for the transverse magnetic (TM) case, for which the magnetic field is out of plane i.e. parallel to the $z$-axis, see Table \ref{tab:physics}).
In the TM case, the parameters $\sigma_\delta$ (resp.
$\Sigma_0$) and $\eta_\delta$ (resp.
$\eta_0$) are respectively the inverse of relative permittivity, and the relative permeability (the latter is assumed to be $1$ if there is no magnetic activity) in the eigenvalue problem \ref{eq:Peb} (resp.
\ref{eq:limitproblem}), which amounts to looking for eigenfrequencies $\Lambda_\delta=H_z(x,y)$ (resp.
$\Phi_0(x,y)=H_z(x,y)$) in the closed cavity $\Pi$, and the Neumann datum holds on $\partial\Pi$ for microwave frequencies i.e.
in the GHz regime (model of a perfect infinite conducting metal for a cavity that is a few centimers in size). One may use results shown in Figures S1-S4 for such a microwave experiment by assuming $c\sim 3 \times 10^{8}\;ms^{-1}$ i.e. the speed of light in vacuum, by considering $cm$ instead of $m$ and $GHz$ instead of $Hz$. Importantly, dielectric layers are non dispersive in the microwave regime and required contrast of their permittivity as shown in Table II is less than 10 and so available. 
There is no obvious way forward in the transverse electric (TE) case as artificial anisotropic permeability is present in the heterogeneous Laplacian operator in \ref{maxcavity1}.
One could possibly achieve some approximate twinning in the TE case (wherein a Dirichlet datum holds) by breaking down the heterogeneous Laplacian into a homogeneous Laplacian plus a small perturbation that would not affect the spectrum of the reference cavity too much.
Similar physical setups hold for anti-plane shear elasticity (in which case the Dirichlet condition amounts to clamped boundary and Neumamnn condition to stress-free boundary), and for pressure acoustic ase (in which case the Neumann boundary condition amounts to a rigid boundary).
We refer to Table \ref{tab:physics} for a summary of the physical settings.
We note that in the case of dielectric waveguides and of three-dimensional cavities, the decoupling between TE and TM polarizations is no longer legitimate and one needs to study the spectrum of the full vectorial Maxwell operator, which is fortunately invariant under geometric transforms see e.g.
\cite{zolla2005foundations}.
Moreover, in finite element models of closed electromagnetic waveguides and three-dimensional cavities, so-called edge-elements are recommended so as to prevent appearance of spurious modes in the spectrum \cite{bossavit1990solving,zolla2005foundations}.

\begin{table}
    \centering
    \begin{tabular}{ccc}
Physics setting & $\sigma$ & $\eta$ \\
SH ($\phi=U_z$) & shear modulus    & mass density \\[.5cm]
TE ($\phi=E_z$) & inverse permeability    & permittivity \\[.5cm]
TM ($\phi=H_z$) & inverse permittivity    & permeability \\[.5cm]
Pressure ($\phi=p$) & inverse density    & inverse bulk modulus
\\[.5cm]
    \end{tabular}
    \caption{Meaning of parameters $\sigma$ and $\eta$ in Eq. (3) and (4) of Letter, for physical settings of anti-plane shear elasticity (SH), wherein $U_z$ is the anti-plane component of the displacement field orthogonal of the $uv$-plane, of transverse electric case (TE), wherein $E_z$ is the longitudinal electric field orthogonal of the $uv$-plane, of transverse magnetic case (TM), wherein $H_z$ is the longitudinal magnetic field orthogonal of the $uv$-plane and acoustic case wherein $p$ is the pressure field.}
    \label{tab:physics}
\end{table}

We further note that the case of a dielectric cavity is more demanding as one would then deal with an unbounded outer medium: theoretically an open cavity has a spectrum that is no longer a purely discrete set of real eigenvalues, and numerically one needs to implement perfectly matched layers \cite{berenger1996perfectly} in the finite element model to look for leaky modes that diverge at infinity. Finally, while it seems natural to also extend the present study that covers the case of anti-plane shear elasticity (see Table \ref{tab:physics}) to the case of in-plane elasticity in two-dimensional cavities, and even to three-dimensional elastic cavities, a word of caution would be in order as the Navier equations are generally not form invariant \cite{milton2006cloaking}.

\section{Acknowledgment}
The numerical simulations for the supplementary material were carried out using the computational facilities of the Advanced Computing Research Centre, University of Bristol - \href{http://www.bristol.ac.uk/acrc/}{http://www.bristol.ac.uk/acrc/}.

\bibliography{bibliography}